\documentclass[11pt,a4paper]{article}
\usepackage{jheppub,amsfonts}
\usepackage[caption=false]{subfig}
\usepackage[titletoc,title]{appendix}
\usepackage{multirow}
\usepackage{longtable}
\usepackage{alphalph}
\usepackage[frozencache=true,cachedir=minted-cache]{minted}
\usepackage{fancyhdr}
\usepackage{tikz}
\usepackage{stackengine}

\newcommand{\beq}{\begin{eqnarray}}
\newcommand{\eeq}{\end{eqnarray}}
\newcommand{\non}{\nonumber\\}

\newcommand{\p}{\partial}
\newcommand{\Tr}{\qopname\relax o{Tr}}
\newcommand{\diag}{\qopname\relax o{diag}}

\newcommand{\btau}{\boldsymbol{\tau}}

\newcommand{\bx}{{\mathbf{x}}}

\newcommand{\bn}{{\mathbf{n}}}
\newcommand{\bR}{{\mathbf{R}}}

\newcommand{\Lag}{\mathcal{L}}

\newcommand{\SU}{\qopname\relax o{SU}}
\newcommand{\SO}{\qopname\relax o{SO}}

\renewcommand{\i}{\mathrm{i}}
\renewcommand{\d}{\mathrm{d}}

\newcounter{Bnum}
\newcounter{sk}[Bnum]
\renewcommand*{\thesk}{\alphalph{\value{sk}}}
\newcommand{\skfig}{%
  \refstepcounter{sk}%
  \thesk%
}

\newcommand{\skfigref}[2]{\hyperref[#2]{$#1_{\rm\ref*{#2}}$}}

\makeatletter
\def\@cline#1-#2\@nil{%
  \omit
  \@multicnt#1%
  \advance\@multispan\m@ne
  \ifnum\@multicnt=\@ne\@firstofone{&\omit}\fi
  \@multicnt#2%
  \advance\@multicnt-#1%
  \advance\@multispan\@ne
  \leaders\hrule\@height\arrayrulewidth\hfill
  \cr
  \noalign{\nobreak\vskip-\arrayrulewidth}}
\makeatother

\newcommand{\numsols}{409}
\newcommand{\newsols}{383}

\title{A Sm\"org\aa sbord of Skyrmions}

\author{Sven Bjarke Gudnason$^1$,}
\affiliation{$^1$Institute of Contemporary Mathematics, School of
  Mathematics and Statistics, Henan University, Kaifeng, Henan 475004,
  P.~R.~China}
\emailAdd{gudnason(at)henu.edu.cn}
\author{Chris Halcrow$^2$}
\affiliation{$^2$School of Mathematics, University of Leeds, 
  Woodhouse Lane, Leeds, LS2 9JT, United Kingdom}
\emailAdd{c.j.halcrow(at)leeds.ac.uk}

\abstract{
We study static solutions of the standard Skyrme model with a pion mass.
Using approximately $10^5$ pseudo-random initial configurations made of
single Skyrmions in the non-symmetrized product Ansatz and an
automatic detection of repeated solutions, we find \numsols{} local
energy minimizers (Skyrmions) of the model with baryon numbers 1
through 16, of which \newsols{} are new. 
In particular, we find new solutions for baryon numbers 5, 8,
9, 10, 11, 12, 13, 14, 15, and 16.
Our results for the number of solutions per baryon number suggest that
this number could grow either polynomially or exponentially. 
We identify new families of solutions: sheets of Skyrmions
in synchronized and antisynchronized hexagonal layers (which we call
graphene); chains of 2- and 3-tori; chain-like solutions containing a
hinge and many clustered Skyrmions. 
Contrary to common lore, only the $B=12$ global energy minimizer
is made of alpha particles or some chunk of a cubic crystal, whereas
the $B=9,11,14,15$ minimizers contain the $B=7$ icosahedrally
symmetric Skyrmion as a component. The $B=10,13,16$ are symmetric
graphene-like solutions. We find $B=5$ and $B=8$ minimizers with
numerically indistinguishable energies.
The $B=8$ candidates are the chain of two cubes,
which is a chunk of the cubic Skyrme crystal and the fullerene-type
ball found originally by the rational map approximation. The $B=5$ 
global minimizer is either the well-known $D_{2d}$ symmetric fullerene
or a new $C_{2v}$ symmetric solution. 
Finally, our findings show a large number of solutions have no
discrete symmetries or just one symmetry, contrary to the common lore
that Skyrmions are highly symmetric configurations.
}
\dedicated{\bigskip\bigskip
\parbox{10cm}{Sm\"org\aa sbord or sm\o rbr\o dsbord (in Swedish and
Danish, respectively) are composite words made of sm\" or (butter),
g\aa s (goose), bord (table) and br\o d (bread). It denotes the meal
and literally the table filled with a wide variety of dishes and
condiments used for the guests to assemble and eat open-faced rye
bread sandwiches. }}

\begin{document}
\maketitle

\section{Introduction}

In the 1960s, Tony Skyrme proposed an elegant model of nuclei as solitons in a
nonlinear sigma model \cite{Skyrme:1961vq,Skyrme:1962vh}. The model is
consistent with chiral symmetry and contains the (unique) leading
second-order term in the chiral Lagrangian as well as a special
fourth-order term, now known as the Skyrme term.
Nontrivial topology means that field configurations can be labeled
by a conserved topological charge $B\in\mathbb{Z}$, which is
identified with the baryon number. In the basic theory, each classical
energy minimizer (Skyrmion) is identified with a nucleus and a
semiclassical quantization allows for comparison with nuclear
data. Early studies were made giving qualitative results for the
single nucleon matching experimental data at about the 30\%
level \cite{Adkins:1983ya}, see ref.~\cite{Zahed:1986qz} for a review.  

To make the connection with nuclear physics, one must first find the
classical solution for baryon number $B$. The mass, radius, moments of
inertia and symmetries of the solution are needed to probe the
properties of nuclei. Even for the deuteron ($B=2$), the problem is
nontrivial. The shape of this solution was settled after some
discussion and the result is that two single Skyrmions merge into a
torus, yielding a classical solution with axial
symmetry \cite{Kopeliovich:1987bt,Manton:1987xf,Verbaarschot:1987au}.

A multitude of numerical methods were used to determine the Skyrmion
solutions with $B>2$ \cite{Braaten:1989rg, Battye:1997qq,
Battye:2004rw}, and the solutions for $B<8$ were found to be hollow
fullerenes. The analytic breakthrough to understand these solutions
came from the Rational Map Approximation (RMA).
Here, the target 3-sphere of the Skyrme model is decomposed into a
radial direction and a rational map describing the angular
distribution of the solution \cite{Houghton:1997kg}.
This allowed for a reconstruction of the known numerical minimizers
and a detailed study of the Finkelstein-Rubinstein
constraints \cite{Krusch:2005iq}, which tell us how the symmetries
restrict the allowed quantum states of the theory. The RMA predicts
that these hollow fullerenes are the energy minimizers as $B$ grows.

However, these large hollow shells are only the minimizers of the energy
functional when the pion mass term is
excluded \cite{Battye:2004rw,Battye:2006tb}. The solutions drastically
change when the pion mass term is included, since the hollow region inside
the fullerenes correspond to the anti-vacuum of the pion fields (the
antipodal point on the target space 3-sphere) and is penalized by the
pion mass term. Since the surface area of the shell grows like
$r^2$, whereas the volume of the shell grows like $r^3$, small enough
fullerenes are allowed and the expectation is that large fullerene
solutions collapse into smaller clusters.
It was suggested that these clusters would be alpha particles of $B=4$
cubes, which would be a welcome result as it connects the Skyrme model to
the well-known alpha-particle model of nuclei. Low energy solutions
were found with $B=8,12,16$ which look like cubes arranged in
geometric shapes \cite{Battye:2006na}.
The cubes also appear in the infinite crystal solution of the Skyrme
model. Hence it seems reasonable that medium to large Skyrmions are
composed of $B=4$ cubes in a variety of arrangements. Results
supporting this lore
include \cite{Feist:2012ps,Lau:2014baa,Halcrow:2016spb}. In contrast
to this cluster picture, it has also been found that the hollow
fullerenes collapse into flat structures \cite{Battye:2004rw}.
There are also several proposals for Skyrmions with special baryon
numbers which have high symmetry, such as tetrahedral
symmetry \cite{Manton:2017bee,Halcrow:2019myn,Manton:2020mpy,Manton:2021xaw}.
Overall, there are many proposals for the form of large $B$ Skyrmions
with nonzero pion mass. 

The shape and symmetry of the classical solutions is important in the
semiclassical expansion generally used to construct nuclear 
spectra from Skyrmions. In a rigid body quantization, the Skyrmion is
allowed to rotate and isorotate and these symmetries lead to spin and
isospin quantum numbers.
However, the discrete symmetry of the Skyrmion restricts which values
of spin and isospin are allowed about each solution, due to the
Finkelstein-Rubinstein
constraints \cite{Finkelstein:1968hy}. Furthermore, the classical
solutions also contain information about the mass and moments of
inertia, which feed into the energies of the quantum states.
Rigid body quantization relies on several approximations. The
wavefunction is a $\delta$-function on configuration space, which
approximates a harmonic oscillator focused around that minimizer. For
this to be accurate, the classical solution must be isolated from
other solutions: either by a potential energy barrier or a large
distance in the space.
If the energy landscape is relatively flat with many local minima, the
approximation breaks down and another quantization method should be
used. The spins of excited nuclear states are governed by the classical
vibrational modes that the particular Skyrmion
possesses \cite{Gudnason:2018bju}. The small number of vibrational
quantizations which have been attempted rely on similar
approximations to rigid body quantization: that the configurations
which are included in the quantization are separated from other
solutions. 
These approximations all make sense if, for example, there is only one
energy minimizing Skyrmion per baryon number. We will soon see that
this is only true for small $B$.

In this paper, we search for the classical local (and global) energy
minimizing solutions, i.e.~the Skyrmions. We use a state-of-the-art
arrested Newton flow code implemented in \texttt{CUDA C} and run on
an \texttt{NVIDIA} GPU high-performance computing cluster. On average,
a solution takes about 3-8 minutes to complete and we have run an
order of $10^5$ calculations with random initial conditions, described
in more detail in the text. 
In order to sort through the results, we have implemented a code that
can recognize repeated solutions and only flag solutions in the
``gray zone'' for human verification. 
The ``gray zone'' comes about due to the existence of soft modes,
i.e.~paths between two solutions with
an extremely small potential barrier.

We have found \numsols{} Skyrmions with $B=1-16$ in the Skyrme model
with a pion mass, including \newsols{} which are new. This is by far
the largest number of Skyrmions ever found. We find that although the
cube is very stable and a common sub-cluster in large massive
Skyrmions, a slew of other types of solution exist. As well as the
cube, large Skyrmions contain arrangements of the $B=7$ icosahedron as
well as other smaller Skyrmions.
We also find new types of solutions which look like graphene layers
with holes either synchronized or anti-synchronized; we dub this
kind of solution symmetric and out-of-phase graphene with an obvious
abuse of terminology.
We find the well-known chains of cubes, which are part of the
cubic Skyrme crystal. But we also find that chains of $B=7$ Skyrmions
exist and can have a smaller energy than the chains of cubes; for example,
such a chain of 7-Skyrmions is the global minimizer of the energy
functional for $B=14$. 
A new finding is a slight variation of the chain of cubes, which we
call hinge Skyrmions and are bent chains with a 1-, 3-, 5- or 7-
Skyrmion acting as a hinge which allows for the bend. Another
interesting phenomenon is that although several configurations are
unstable by themselves, they are a viable component in larger
Skyrmions. The most common of these is the 3-torus.

The paper is organized as follows. In sec.~\ref{sec:model}, we review
the massive Skyrme model, detail the moment of inertia tensors and translate
the results to vector notation (useful for numerical computations),
review the energy bound for the model including the pion mass term,
define the characteristics used to distinguish solutions and explain
how to identify discrete symmetries possessed by the solutions.
In sec.~\ref{sec:numerical}, we explain the algorithm to generate
pseudo-random initial configurations, the numerical method and the 
method we use to automatically eliminate already-found solutions.
In sec.~\ref{sec:results}, we discuss the results and the new
types of solutions found in this work, as well as present statistics
based on the properties of all \numsols{} solutions. 
Finally, we conclude the paper with a discussion and outlook in
sec.~\ref{sec:discussion}.
Tables containing plots and details of the \numsols{} solutions are
presented in app.~\ref{app:sols}. The solutions are also available
online accompanied by animations of each solution rotating
at \href{http://bjarke.gudnason.net/smoergaasbord}{http://bjarke.gudnason.net/smoergaasbord}. 

\section{The massive Skyrme model}\label{sec:model}

The standard Skyrme model with a pion mass term, is given by
\beq \label{eq:lag}
\Lag = \Tr\left[\frac12L_\mu L^\mu + \frac{1}{16}[L_\mu,L_\nu][L^\mu,L^\nu] -
  m^2(\mathbf{1}_2 - U)\right],
\eeq
where $m$ is the pion mass in Skyrme units, $L_\mu=U^\dag\p_\mu U$
is the left-invariant chiral current, $\mu,\nu=0,1,2,3$ are spacetime
indices and $U$ is the chiral Lagrangian field or simply the Skyrme
field, related to pions by 
\beq
U = \mathbf{1}_2\sigma + \i\pi^a\tau^a, 
\eeq
with $a=1,2,3$; $\sigma$ is an auxiliary field making the nonlinear
sigma model constraint $\det U=1$ possible and we use the
mostly-positive metric convention, useful for static solitons. 

The equation of motion for the Lagrangian \eqref{eq:lag} is
\beq
\p_\mu \left(L^\mu - \frac14\big[L^\nu,[L_\nu,L^\mu]\big] - m^2 \widehat{U}\right) = 0,\qquad
\widehat{U}:= U - \frac12\Tr[U]
\label{eq:eom}
\eeq
and the static energy reads
\beq
E = \int_{\mathbb{R}^3}\Tr\left[
  -\frac12 L_i L_i - \frac{1}{16}[L_i,L_j][L_i,L_j] + m^2(\mathbf{1}_2 - U)
  \right]\d^3x.
\label{eq:E}
\eeq
In this paper we define a Skyrmion to be a solution to the equation of
motion \eqref{eq:eom} which is at least a local minimum of the static
energy. 

Static, finite-energy configurations must satisfy $U\sim \mathbf{1}_2$
as $|x| \to \infty$, causing a one-point compactification of
space. This means that the static Skyrme field is a map from
$S^3 \to S^3$, allowing for nontrivial topology. Each
configuration has a conserved integer topological charge, which is
given by
\beq
B = -\frac{1}{24\pi^2}\int_{\mathbb{R}^3} \epsilon_{ijk}\Tr[L_i L_j L_k]\;\d^3x.
\eeq
The energy of a Skyrmion with charge $B$ is bound by $B$ and grows
approximately linearly with $B$. 

The $B=1$ Skyrmion has a spherically symmetric energy density.
The Skyrme field takes the form
\beq
U = \mathbf{1}_2\cos f(r) +
i\hat{\boldsymbol{x}} \cdot \tau \sin f(r),
\eeq
where $r=|x|$ and $f(r)$ is the profile function.
We can insert this Ansatz into the Lagrangian to find the equation for
$f(r)$, 
\beq \label{eq:prof}
(r^2 + 2\sin^2f)f'' + 2rf'
+\sin 2f\left(f'{}^2 - 1 - \frac{\sin^2 f}{r^2}\right)
-r^2m^2\sin f = 0.
\eeq

Although pions have a fixed physical mass, the dimensionless
value depends on the calibration since $m=\frac{2m_\pi}{eF_\pi}$. As
there are various ways to calibrate the Skyrme model, previous works have
used various $m$, such as $m=0.526$ \cite{Adkins:1983hy}, $m=1$
\cite{Battye:2006tb} and $m=1.125$ \cite{Manko:2007pr}. As we are
interested in qualitative features of the Skyrmions, we choose
$m=1$. This also allows us to compare our results with previous
studies \cite{Battye:2006tb,Battye:2006na}. Our results, especially
the properties of the solutions, would change if $m$ was
changed. There are also known solutions with $m=0$ which are
unstable when $m=1$, and vice versa. Hence we expect Skyrmions to
appear and disappear as $m$ changes.

\subsection{Kinetic energy and moment of inertia tensors}

Generally a static Skyrmion is one of a nine-parameter family of solutions
with the same energy, related by translations, rotations and
isorotations (rotations of the target 3-sphere). We will fix the
center of mass from now on. We can restrict to just these
configurations by applying rotations and isorotations to the static
solution $U_0$, 
\beq \label{eq:Atime}
U = A U_0(D(B)\bx) A^\dag,
\eeq
via the SU(2) matrices $A=A(t)$ and $B=B(t)$.
The map $D$ : $\SU(2)\to\SO(3)$ is given by
\beq
D(B)_{i j} = \frac12\Tr[\tau^i B\tau^j B^\dag].
\eeq
By restricting to these configurations we interpret the Skyrmion as a
generalized rigid body, with a mass and moments of inertia. We can
calculate the moments of inertia by inserting the
Ansatz \eqref{eq:Atime} into the Lagrangian, yielding the kinetic
energy 
\beq
T = \frac12 a_i U_{i j} a_j
-a_i W_{i j} b_j
+\frac12 b_i V_{i j} b_j,
\label{eq:T}
\eeq
where
\beq
a_i = -\i\Tr[\tau^i A^\dag\dot{A}], \qquad
b_i = -\i\Tr[\tau^i \dot{B}B^\dag],
\eeq
and the isospin, mixed and spin moments of inertia tensors are given
by \cite{Manko:2007pr}
\begin{align}
  U_{i j} &= -\int_{\mathbb{R}^3}\Tr\left[T_i T_j + \frac14[L_k,T_i][L_k,T_j]\right]\d^3x,\\
  W_{i j} &= \int_{\mathbb{R}^3}\epsilon_{j l m} x^l\Tr\left[T_i L_m + \frac14[L_k,T_i][L_k,L_m]\right]\d^3x,\\
  V_{i j} &= -\int_{\mathbb{R}^3}\epsilon_{i l m}\epsilon_{j n p} x^l x^n
  \Tr\left[L_m L_p + \frac14[L_k,L_m][L_k,L_p]\right]\d^3x,
\end{align}
where
\beq
T_i = \frac{\i}{2}U^\dag[\tau^i,U],
\eeq
and $\{\tau^i\}$, $i=1,2,3$, are the Pauli spin matrices.

\subsection{Vector formulation}

For numerical purposes, it will prove convenient to rewrite the Skyrme
model and the above equations in terms of the 4-vector field
$\bn=\{n^0,n^1,n^2,n^3\}$,
related to $U$ by
\beq
U = \mathbf{1}_2n^0 + \i n^a\tau^a, \qquad a=1,2,3.
\eeq
The Lagrangian reads
\beq
\Lag = -\p_\mu\bn\cdot\p^\mu\bn
-\frac12(\p_\mu\bn\cdot\p^\mu\bn)^2
+\frac12(\p_\mu\bn\cdot\p_\nu\bn)(\p^\mu\bn\cdot\p^\nu\bn)
-2m^2(1-n^0).
\eeq
The equation of motion is
\begin{align}
&\p^2n^A
+(\p_\nu\bn\cdot\p^\nu\bn)\p^2n^A
+(\p_\mu\p_\nu\bn\cdot\p^\nu\bn)\p^\mu n^A\non
&-(\p^2\bn\cdot\p_\mu\bn)\p^\mu n^A
-(\p_\mu\bn\cdot\p_\nu\bn)\p^\mu\p^\nu n^A
+m^2 \delta^{A0} = 0, \qquad A=0,\ldots,3,
\label{eq:eom_n}
\end{align}
and the static energy reads
\begin{equation} \label{eq:En}
E = \int_{\mathbb{R}^3}\left[
\p_i\bn\cdot\p_i\bn
+\frac12(\p_i\bn\cdot\p_i\bn)^2
-\frac12(\p_i\bn\cdot\p_j\bn)(\p_i\bn\cdot\p_j\bn)
+2m^2(1-n^0)
\right]\d^3x.
\end{equation}
The topological charge is
\beq
B = \frac{1}{2\pi^2}\int_{\mathbb{R}^3}\epsilon^{A B C D}n^A\p_1n^B\p_2n^C\p_3n^D, \qquad
A,B,C,D=0,\ldots,3,
\eeq
where we use the convention $\epsilon^{0123}=1$.

The kinetic energy is still given by eq.~\eqref{eq:T}, but with the
tensors
\begin{align}
  U_{i j} &= 2\int_{\mathbb{R}^3}\Big[
    n^a n^a\delta_{i j}
    -n^i n^j
    +(\delta_{i j} - n^i n^j)(\p_k n^0)^2
    +n^a n^a \p_k n^i\p_k n^j\non
    &\phantom{=2\int_{\mathbb{R}^3}\Big[\ }
    +n^0 n^i\p_k n^0 \p_k n^j
    +n^0 n^j\p_k n^0 \p_k n^i\Big]\d^3x,\\
    W_{i j} &= 2\int_{\mathbb{R}^3}\epsilon_{j l m}x^l\epsilon^{i a b} n^a \left[
      \p_m n^b
      +\p_m n^b\p_k\bn\cdot\p_k\bn
      -\p_k n^b\p_k\bn\cdot\p_m\bn
      \right]\d^3x,
    \\
    V_{i j} &= 2\int_{\mathbb{R}^3}\epsilon_{i l m}\epsilon_{j n p}x^lx^n\Big[
      \p_m\bn\cdot\p_p\bn
      +(\p_m\bn\cdot\p_p\bn)(\p_k\bn\cdot\p_k\bn)\non
    &\phantom{=2\int_{\mathbb{R}^3}\Big[\ }
      -(\p_m\bn\cdot\p_k\bn)(\p_p\bn\cdot\p_k\bn)\Big]\d^3x,\label{eq:V}
\end{align}
with $i,j,k,l,m,n,p,a=1,2,3$. 

\subsection{Energy bound}

In many past studies, the energy of a Skyrmion has been presented in
units of $E/12\pi^2$ following Skyrme's energy bound
$E\geq12\pi^2$. Hence, the energy presented displays how close the
solution is to saturating the bound. Since then, Skyrme's original
bound has been improved when the pion mass is included. As we are
using the pion mass, we will use an updated bound. Following
refs.~\cite{Harland:2013rxa,Adam:2013tga}, we can write the 
topological energy bound for the energy \eqref{eq:E} (or
\eqref{eq:En}) as
\begin{align}
  E &= E_2 + E_4 + 2m^2E_0\\
  &= (E_2 + \alpha E_4) + \big(2m^2 E_0 + (1-\alpha)E_4\big)\label{eq:partition}\\
  &\geq 12\pi^2\left[\sqrt{\alpha} + \frac{128\sqrt{m}(1-\alpha)^{\frac34}\Gamma^2\big(\tfrac34\big)}{45\pi^{\frac32}}\right]|B|, \qquad
  \alpha\in[0,1],
  \label{eq:alpha_bound}
\end{align}
where we have used the subbounds
\begin{align}
  \beta E_2 + E_4 &\geq 12\pi^2\beta^{\frac12}|B|,\\
  \beta E_0 + E_4 &\geq 8\pi^2\beta^{\frac14}\big\langle\mathcal{E}_0^{\frac14}\big\rangle|B|,
\end{align}
and we have defined
\begin{align}
  E_2 &= -\frac12\int_{\mathbb{R}^3}\Tr[L_i L_i]\;\d^3x
  = \int_{\mathbb{R}^3} \p_i\bn\cdot\p_i\bn\;\d^3x,\\
  E_4 &= -\frac{1}{16}\int_{\mathbb{R}^3}\Tr[L_i,L_j][L_i,L_j]\;\d^3x
  = \frac12\int_{\mathbb{R}^3} \left[(\p_i\bn\cdot\p_i\bn)^2 - (\p_i\bn\cdot\p_j\bn)(\p_i\bn\cdot\p_j\bn)\right]\;\d^3x,\\
  E_0 &= \int_{\mathbb{R}^3}\mathcal{E}_0\d^3x
  = \frac12\int_{\mathbb{R}^3}\Tr[\mathbf{1}_2 - U]\;\d^3x,
\end{align}
as well as the target space average
\beq
\big\langle X\big\rangle = -\frac{1}{24\pi^2B}\int_{\mathbb{R}^3}
X \epsilon_{i j k}\Tr[L_i L_j L_k]\;\d^3x.
\eeq
We obtain
\begin{align}
  \big\langle1\big\rangle &= \frac{2}{\pi}\int_0^\pi\sin^2\xi\;\d\xi = 1,\\
  \big\langle\mathcal{E}_0\big\rangle &= \frac{2}{\pi}\int_0^\pi\sin^2\xi(1-\cos\xi)\;\d\xi = 1,\\
  \big\langle\mathcal{E}_0^{\frac14}\big\rangle
  &= \frac{2}{\pi}\int_0^\pi\sin^2\xi(1-\cos\xi)^{\frac14}\;\d\xi =
  \frac{32\Gamma^2\big(\tfrac34\big)2^{\frac34}}{15\pi^{\frac32}}.
\end{align}
In order to obtain the strongest possible bound from
eq.~\eqref{eq:alpha_bound}, one must maximize the bound in terms of
$\alpha$.
The solution to
\beq
\frac{\;\d}{\d\alpha}\left(\sqrt{\alpha} + \frac{2}{3\sqrt{a}}(1-\alpha)^{\frac34}\right) = 0,
\eeq
is given by
\beq
\alpha = \frac{a^2}{2}\left(\sqrt{1 + \frac{4}{a^2}} - 1\right), 
\eeq
and from eq.~\eqref{eq:alpha_bound}, we can read off $a$ as
\beq
a = \frac{225\pi^3}{4096m\Gamma^4\big(\tfrac34\big)}.
\eeq
Setting $m:=1$, we obtain the following numerical values
\begin{align}
  a \simeq 0.75533, \qquad
  \alpha \simeq 0.52214, \qquad
  E \gtrsim 12\pi^2\times 1.16347.
  \label{eq:bound_num}
\end{align}
One can also check that if $m:=0$, the bound
\eqref{eq:alpha_bound} is maximized by $\alpha=1$ and is hence the
result found by Skyrme \cite{Skyrme:1961vq}, i.e.~$E\geq12\pi^2$.
Interestingly, we can see that when $m=1$, $\alpha\approx\tfrac12$. This
means that the Skyrme term is shared almost equally between the two
partitions in eq.~\eqref{eq:partition}, and hence the energy bound
involving the pion mass term has a significant impact on the total
energy bound.

\subsection{Characteristics}\label{sec:characteristics}

In order to distinguish between Skyrmion solutions we need to define some
characteristics of the solutions.

First, we can calculate the energy \eqref{eq:En} divided by the
topological energy bound \eqref{eq:bound_num}:
\beq
\epsilon \equiv \frac{E}{12\pi^2GB}, \qquad
G \simeq 1.16347,
\label{eq:epsilon_def}
\eeq
which tells us how much the energy of a given solution is above the
bound.
For convenience, we have also divided by the baryon number 
(topological degree), so it is easy to compare the energy between
different topological sectors. 

Second, we can calculate the radius of the Skyrmion, using here the
topological charge density as argued for in ref.~\cite{Adam:2015zhc},
\beq
\mathfrak{r} \equiv
\sqrt{\frac{1}{2\pi^2B}\int_{\mathbb{R}^3}(\bx-\bx_0)\cdot(\bx-\bx_0)\,
  \epsilon^{A B C D} n^A\p_1 n^B\p_2 n^C\p_3 n^D\;\d^3x},
\eeq
where $\bx$ is the coordinates in $\mathbb{R}^3$ and $\bx_0$ is the
center of charge density
\beq
\bx_0 \equiv
\frac{1}{2\pi^2B}\int_{\mathbb{R}^3}\bx\,\epsilon^{A B C D} n^A\p_1 n^B\p_2 n^C\p_3 n^D\;\d^3x.
\eeq
The radius $\mathfrak{r}$ is a scalar quantity and is indeed invariant
under rotations.

Third, we can calculate the eigenvalues of the spin and isospin moment
of inertia tensors
\begin{align}
  V_R^{-1} V V_R &= V_D,\label{eq:VD}\\
  U_R^{-1} U U_R &= U_D,
\end{align}
where $V, U$ are the moment of inertia tensors for rotations and
isorotations, $V_R,U_R$ are $\SO(3)$ diagonalization matrices and
$V_D,U_D$ are the eigenvalues (diagonal matrices) of the moment of
inertia tensors.

Fourth, we can identify the symmetry group of the Skyrmion solution,
as we will discuss in the next subsection.

Finally, we will also calculate numerically the amount of the
topological charge density captured by the numerical solution
\beq
b \equiv \frac{1}{2\pi^2B}\int_{\mathbb{R}^3}\epsilon^{A B C D}
n^A\p_1 n^B\p_2 n^C\p_3 n^D\;\d^3x,
\eeq
which mathematically is equal to one.
The deviation from unity is a measure of how precise the numerical
solutions are.

\subsection{Symmetries}

We classify each numerically generated Skyrmion by its symmetry
group. Since large symmetry groups are relatively rare and easy to see
by eye, we only need to consider small groups systematically.
There are eight groups whose generators all have order two or
less. The groups $D_{2d}$ and 
$D_{2h}$ have order 8, $D_2$, $C_{2v}$ and $C_{2h}$ have order 4 while
$C_2$, $C_i$ and $C_{1h}$ have order 2. These eight groups can mostly be
distinguished by the reflection symmetries of a solution.

\begin{figure}[!ht]
\centering
\includegraphics[width=1.0\textwidth]{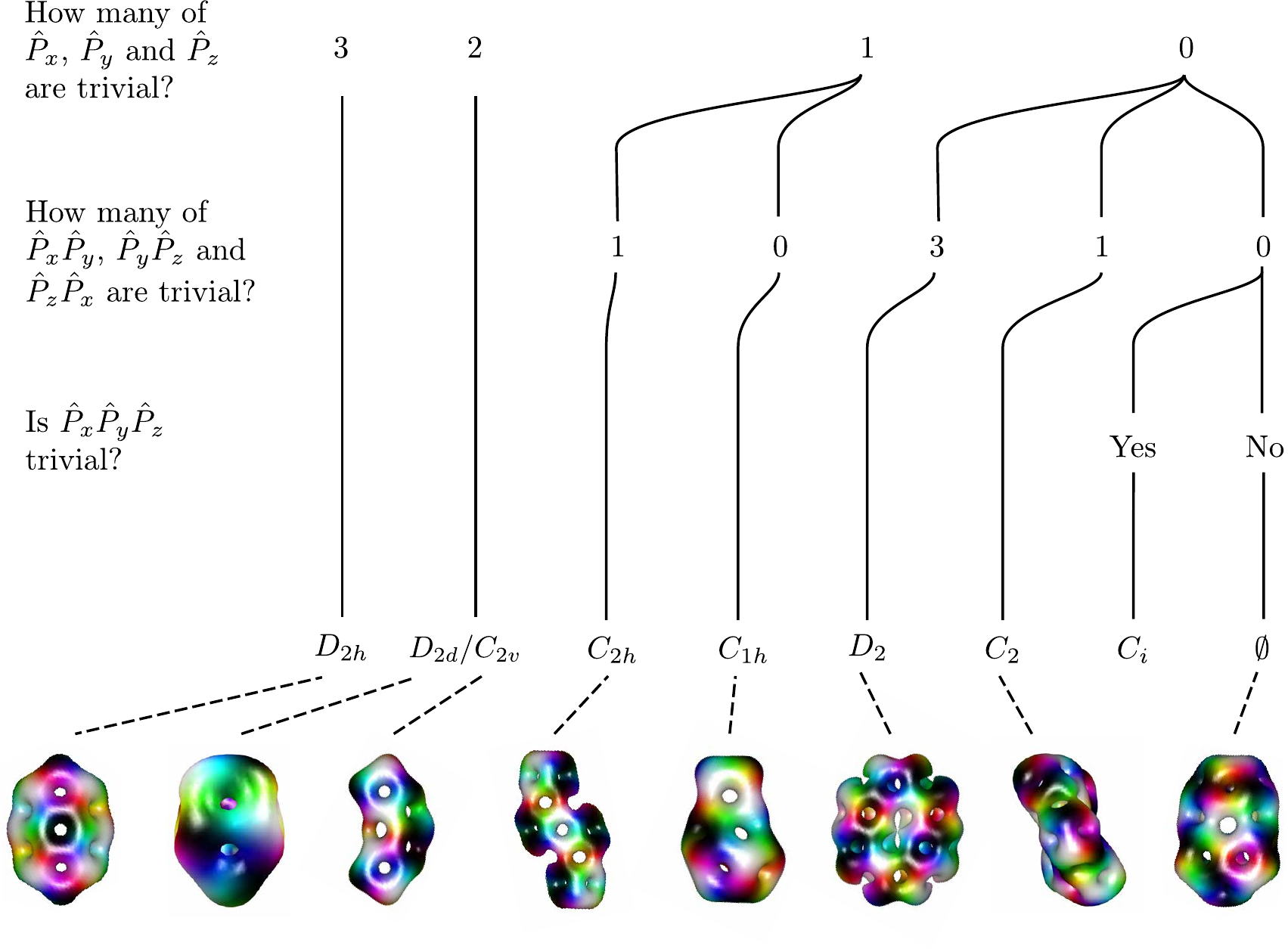}
\caption{A procedure to calculate the symmetry group of a given
solution. An example of a Skyrmion with that symmetry group is given,
where available.}
\label{fig:symmetries}
\end{figure}

If a rigid body has a reflection plane, the normal of this plane will
be a principal axis. Hence if we orient our solution so that the
principal axes match the Cartesian axes, we can determine the
reflection symmetries of the Skyrmion by applying reflection operators
to the energy density and checking if they are trivial or not. Denote
the reflection operators as $\hat{P}_x: x \to -x$ etc. Then, to
calculate the symmetry of the solution we follow the procedure
displayed in fig.~\ref{fig:symmetries}. The one awkward case is
differentiating between $D_{2d}$ and $C_{2v}$ symmetry. The $D_{2d}$
group does not have a unique principal axes. We choose the axes equal
to the Cartesian axes and with this choice, the two groups cannot be
distinguished by reflections alone. We check these cases by eye: the
$D_{2d}$ group has a distinctive $90^\circ$ rotation plus reflection
symmetry. 

For most symmetry groups, we find a Skyrmion with that symmetry. The
only exception is the $C_i$ group: a group of order two where the only
symmetry is a full reflection $(x,y,z) \to (-x,-y,-z)$.

\section{Numerical computations}\label{sec:numerical}

\subsection{Initial configurations}

In order to find as many possible global and local minima of the
energy functional we generate initial conditions using a
standard pseudo-random generator in \texttt{C}.
Since we only generate a single configuration per numerical
computation, we use a CPU code to generate the random initial
parameters, but a GPU code to generate the initial data on the lattice. 

The algorithm we use to generate a random initial configuration is
described as follows. We first generate a 3-tuple of random numbers
for the position and a 3-tuple of random numbers for the 3 rotational Euler
angles of a 1-Skyrmion (rotation in space and in isospace are
equivalent for a 1-Skyrmion); from now on we will call this a 6-tuple
of moduli of the 1-Skyrmion. Then for each additional 1-Skyrmion (that
is for the $B-1$ remaining 1-Skyrmions), we generate a new 6-tuple of 
random numbers. The new 1-Skyrmion is only accepted if the distance
between it and one of the previously generated 1-Skyrmions is smaller than a
fixed length. This criterion avoids the possibility of having isolated
Skyrmions of topological degrees $B_1$ and $B_2$ at different places
in space. After having fixed all the moduli randomly for each of the
$B$ 1-Skyrmions, we move the center of mass to the origin of
space. Finally, we will create the initial configuration using the
asymmetric product Ansatz $U=U_1U_2\cdots U_B$, which preserves the
nonlinear sigma model constraint $\det U=1$ and
\beq
U_i = \mathbf{1}_2\cos f(r_i) +
\frac{\i(\bR_i(\bx-\bx_{i}))\cdot\btau}{r_i}\sin f(r_i), \qquad i=1,2,\ldots,B,
\eeq
with $r_i=\sqrt{(\bx-\bx_{i})\cdot(\bx-\bx_{i})}$ the distance
from the position $\bx_{i}$, $\bR_i$ a rotation matrix and
$f(r_i)$ the profile function of the 1-Skyrmion, which is a solution
to the equation \eqref{eq:prof}.

The algorithm in pseudo code is:
\begin{minted}[fontsize=\footnotesize]{matlab}
  x[0] = genRand3(XMIN,XMAX);
  alpha[0] = genRand3(0,2*pi);
  for (i=1:B-1)
    minlen = XMAX;
    do
      x[i] = genRand3(XMIN,XMAX);
      for (j=0:i-1) if (vectorlength(x[i]-x[j]) < minlen) minlen = vectorlength(x[i]-x[j]);
    while (minlen > MAXSEPARATION);
    alpha[i] = genRand3(0,2*pi)  
  end for
  com = {0,0,0};
  for (i=0:B-1) com += x[i];
  com /= B;
  for (i=0:B-1) x[i] -= com;
  create1Skyrmion(U,x[0],alpha[0])
  for (i=1:B-1) add1SkyrmionByRightMultiplication(U,x[i],alpha[i]);
\end{minted}
We have defined suitable functions, constants are written in
caps and the latter two lines are implemented as GPU code.
We have tried several values for MAXSEPARATION and have mostly used
MAXSEPARATION\ $=2$.

\subsection{Evolution}

We approximately solve the equation of motion \eqref{eq:eom_n} by
including the time-dependent part of the kinetic term ($\p_t^2n^A$)
but not the time-dependent parts of the higher-order derivative
terms. The initial configuration is not a minimizer and so as it
evolves it trades potential energy for kinetic energy. We monitor the
static potential energy $E$ \eqref{eq:En}, and eliminate the entire 
kinetic energy by setting $\p_tn^A=0$ whenever $E$ increases during a
computation. After several iterations the configuration will lie at
the bottom of a potential well, and is a local or global energy
minimizer.
This method is sometimes called the arrested Newton flow
\cite{Gudnason:2020arj}.
We stop the evolution when $\sum_{\textrm{lattice
    sites}}\sum_{A=1}^4|\frac{\delta E}{\delta
  n^A}|^2<\epsilon_{\text{th}}$ for some threshold
$\epsilon_{\text{th}}<0.01$.

We solve the differential equations by using the finite difference
method with a 4th-order discretization of the partial derivatives with
a 5-point stencil on a cubic lattice with $121^3$ lattice sites with
lattice spacing $0.1333$.
Since we perform a large number of numerical computations, we
have written the numerical code in \texttt{CUDA C} for \texttt{NVIDIA}
GPUs and we run our code on a GPU  high-performance computing (HPC)
cluster. Once we have found the final energy minimizers, we repeat the
calculation on a $361^3$ lattice with lattice spacing $0.0444$, to
increase the accuracy of the final energy. 
We estimate the systematic error by integrating the topological charge
density and seeing how close it is to its integer value. This estimate is
only reliable because we have tested that the threshold for stopping
the computations is good enough that it does not give a significant
contribution, except for soft modes where one should bear in mind that
the numerical error may be larger than reported in the appendix.

It is well known that 1-Skyrmions attract or repel depending on their
initial relative orientation. Despite this, we find that the initial
Skyrmions generally attract, given enough time. Occasionally, a
1-Skyrmion is repelled by the others. In this case, the 1-Skyrmion
moves away from the others and our boundary conditions allow for it
to leave the box. At the end of the evolution we calculate the total
baryon number $B$, which can differ from the initial baryon number
$B_{\rm in}$ due to this phenomena. The final solution is considered to
be a $B$-skyrmion.

\subsection{Elimination of doublets}\label{sec:elimination}

Each run of our code starts with an initial configuration and generates
a Skyrmion. Although every initial condition is random and most likely
different from the previous ones, the Skyrmion is often the same as
one found in a previous computation, perhaps in a different (iso)orientation.
This is to be expected, since all initial configurations within a
Skyrmion's basin of attraction should evolve into that Skyrmion under
arrested Newton flow. 

In order to eliminate the doublets (or $N$-plets) of the same
 Skyrmion, we have to characterize the solution in terms of a set of
numbers that can be compared by a computer algorithm so as to
determine whether a newly found solution is new or already known.
We use the eight characteristic parameters defined in
sec.~\ref{sec:characteristics}, i.e.~the energy normalized by its
topological energy bound $\epsilon$, the radius $\mathfrak{r}$ and the
three eigenvalues of the moment of inertia tensor $V_D$ as well as the
three eigenvalues of the isospin moment of inertia tensor $U_D$.

We define a simple distance in solution space by
\beq
\mathfrak{d}
=
\left(
\left|\frac{\epsilon-\epsilon'}{\min(\epsilon,\epsilon')}\right|,
\left|\frac{\mathfrak{r}-\mathfrak{r}'}{\min(\mathfrak{r},\mathfrak{r}')}\right|,
\left|\frac{V_D-V_D'}{\min(V_D,V_D')}\right|,
\left|\frac{U_D-U_D'}{\min(U_D,U_D')}\right|
\right),
\eeq
where the latter two expressions are 3-vectors, and hence are
evaluated for each of the eigenvalues, and the primed and unprimed
quantities are for the newly found trial solution and one of the known
Skyrmion solutions respectively.

So that we do not calculate the Skyrmion characteristics every time a
comparison is carried out, we save the characteristics of the known
Skyrmions in a log file. Every time a new Skyrmion candidate is found,
its characteristics are computed and the vector $\mathfrak{d}$ is
calculated for each solution in the database (with the same
topological charge).
The new candidate solution is rejected if
$\mathfrak{d} < \mathfrak{d}_{\rm threshold}$. Practical tests reveal
that a suitable threshold distance in solution 
space is $\mathfrak{d}_{\rm threshold}=0.004$. If all $\mathfrak{d}$
are smaller than $\mathfrak{d}_{\rm threshold}$ then the solution is
already known; if one or more are larger then it is a new Skyrmion
solution.
We define a ``gray zone'' for $\mathfrak{d}\in[0.004,0.008]$, which is
then checked manually -- we discuss this in more detail in
sec.~\ref{sec:softmodes}.

\subsection{Visualization}
We visualize Skyrmions by first plotting an isosurface of constant
baryon charge density. We then color the isosurface depending on the
value of the pion field at that point, based on the Runge color
sphere. We color the Skyrmion white/black when $\pi_3 = \pm1$ and red,
green and blue when $\pi_1+\i\pi_2$ is equal to 1, $\exp(2\pi\i/3)$ and
$\exp(4\pi\i/3)$, respectively. Plots of Skyrmions can be found e.g.~in
Fig.~\ref{fig:symmetries}.

\section{Results}\label{sec:results}

\subsection{Minimizers}

\begin{figure}[!htp]
\begin{center}
\mbox{
\stackunder[5pt]{\includegraphics[width=0.095\linewidth]{{B1_121s8c1_crop}}}{\skfigref{1}{skfig:B1c1}}
\stackunder[5pt]{\includegraphics[width=0.095\linewidth]{{B2_121s8c1_crop}}}{\skfigref{2}{skfig:B2c1}}
\stackunder[5pt]{\includegraphics[width=0.095\linewidth]{{B3_121s8c1_crop}}}{\skfigref{3}{skfig:B3c1}}
\stackunder[5pt]{\includegraphics[width=0.095\linewidth]{{B4_121s8c2_crop}}}{\skfigref{4}{skfig:B4c2}}
\stackunder[5pt]{\includegraphics[width=0.095\linewidth]{{B5_121s8c1_crop}}}{\skfigref{5}{skfig:B5c1}}
\stackunder[5pt]{\includegraphics[width=0.095\linewidth]{{B5_121s8c5_crop}}}{\skfigref{5}{skfig:B5c5}}
\stackunder[5pt]{\includegraphics[width=0.095\linewidth]{{B6_121s8c1_crop}}}{\skfigref{6}{skfig:B6c1}}
\stackunder[5pt]{\includegraphics[width=0.095\linewidth]{{B7_121s8c1_crop}}}{\skfigref{7}{skfig:B7c1}}
\stackunder[5pt]{\includegraphics[width=0.1\linewidth]{{B8_121s8c1_crop}}}{\skfigref{8}{skfig:B8c1}}
}
\mbox{
\stackunder[5pt]{\includegraphics[width=0.095\linewidth]{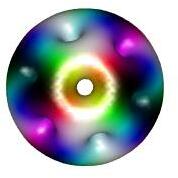}}{\skfigref{8}{skfig:B8c3}}
\stackunder[5pt]{\includegraphics[width=0.095\linewidth]{{B9_121s8c28_crop}}}{\skfigref{9}{skfig:B9c28}}
\stackunder[5pt]{\includegraphics[width=0.075\linewidth]{{B10_121s8c1_crop}}}{\skfigref{10}{skfig:B10c1}}
\stackunder[5pt]{\includegraphics[width=0.09\linewidth]{{B11_121s8c6_crop}}}{\skfigref{11}{skfig:B11c6}}
\stackunder[5pt]{\includegraphics[width=0.14\linewidth]{{B12_121s8c51_crop}}}{\skfigref{12}{skfig:B12c51}}
\stackunder[5pt]{\includegraphics[width=0.07\linewidth]{{B13_121s8c26_crop}}}{\skfigref{13}{skfig:B13c26}}
\stackunder[5pt]{\includegraphics[width=0.1\linewidth]{{B14_121s8c46_crop}}}{\skfigref{14}{skfig:B14c46}}
\stackunder[5pt]{\includegraphics[width=0.15\linewidth]{{B15_121s8c6_crop}}}{\skfigref{15}{skfig:B15c6}}
\stackunder[5pt]{\includegraphics[width=0.08\linewidth]{{B16_121s8c4_crop}}}{\skfigref{16}{skfig:B16c4}}
}
\caption{Minimizers of the Skyrme model with pion mass $m=1$. The
appearance of more than one Skyrmion with same baryon number indicates
close lying states, with increasing energy as the alphabet increases,
but whose energies are within the estimated systematic error. }
\label{fig:minimizers}
\end{center}
\end{figure}

We now present the results of our large scale computations. The global
energy minimizers are shown in fig.~\ref{fig:minimizers}, where we
have included all solutions that are within our total estimated
systematic error, $0.00011$, of the normalized energy. 

The Skyrmions with baryon number 1, 2, 3, 4, 6 and 7 are the
well-known minimizers of the massive (and massless) Skyrme model with
$O(3)$, $D_{\infty h}$, $T_d$, $O_h$, $D_{4d}$ and $Y_h$ symmetry
respectively.
For $B=8$, our code finds two configurations which are too close in
energy to distinguish. These are the chain made of two cubes with
$D_{4h}$ symmetry and the fullerene-type ball with $D_{6d}$
symmetry. Both solutions are well known and our study confirms that
their energies are near degenerate when $m=1$. The surprising result is
that the $B=5$ sector also has two energetically indistinguishable
solutions: one is the known solution with $D_{2d}$ symmetry, the other
has only $C_{2v}$ symmetry. In one orientation this new Skyrmion
looks rather flat, with four holes on top and four on the bottom. The
holes are out of phase and it could be interpreted as a
deformed version of the previously studied $D_{4d}$-symmetric
5-Skyrmion \cite{Manko:2007pr}. The 
$D_{2d}$-symmetric $B=5$ has one anomalously low vibrational
frequency \cite{Gudnason:2018bju}: perhaps the two solutions are
related through this mode.

The minimizer for $B=9$ is a \skfigref{7}{skfig:B7c1}-Skyrmion with a
2-torus attached to a face. The minimizer for $B=10$ is of a type we
will denote ``symmetric graphene'' (see next subsection) and has
been previously found by Battye and
Sutcliffe \cite{Battye:2006tb}. The $B=13$ and $B=16$ minimizers are
similar to that of the $B=10$ sector, extended in a natural way.
For $B=11$, the minimizer is a \skfigref{7}{skfig:B7c1}-Skyrmion with
a cube attached in such a way that there is a retained reflection
symmetry.
The $B=12$ minimizer is as expected from previous work: a chain of
cubes with the middle cube rotated by 90 degrees around the axis
joining them.
The minimizer for $B=14$ is made of two \skfigref{7}{skfig:B7c1}-Skyrmions.
The minimizer for $B=15$ is the \skfigref{7}{skfig:B7c1}-Skyrmion
attached to the end of the \skfigref{8}{skfig:B8c1}-Skyrmion.

These minimizers show that the icosahedral $B=7$ Skyrmion plays an
important role as a cluster in low energy configurations, as well as
the cubic $B=4$ Skyrmion. There are many $\alpha$-cluster models of
nuclear physics but no $^7$Li-cluster models. This may be because of
the fermionic nature of Lithium-7: it must have a quantum spin
contribution, meaning that it is more difficult to use as a 
building block than the spinless $\alpha$-particle.  Classically,
there is no such restriction.

\subsection{Types of solution}

Although we have found many new Skyrmions, they can mostly be sorted
into a small number of different families of solutions. In this
section we will explore these families, with examples.

\begin{figure}[!h]
\centering
\begin{tikzpicture}
\node[inner sep=0] (image) at (0,0) {\includegraphics[width=0.9\textwidth]{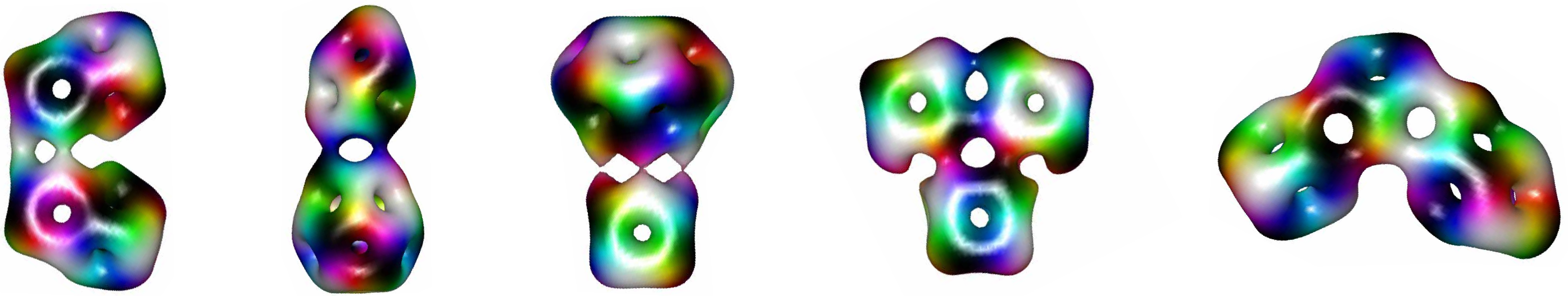}};
\node at (-6.9,-1.2) {\skfigref{12}{skfig:B12c73}};
\node at (-4.5,-1.2) {\skfigref{12}{skfig:B12c61}};
\node at (-2,-1.2) {\skfigref{12}{skfig:B12c60}};
\node at (0.9,-1.2) {\skfigref{12}{skfig:B12c13}};
\node at (3.9,-1.2) {\skfigref{12}{skfig:B12c15}};
\end{tikzpicture}
\caption{Some $B=12$ Skyrmions made from smaller clusters.}
\label{fig:12_clusters}
\end{figure}
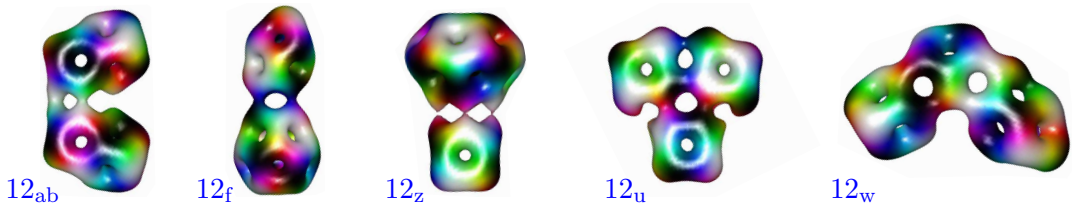

For small $B$, the Skyrmions form symmetric polyhedra. Most of the
small-$B$ solutions ($B<8$) are known, although we have found a new
$B=5$ solution with $C_{2v}$ symmetry. Many large-$B$ Skyrmions look
like loosely bound clusters of these 
lower-charge building blocks. All small $B$ Skyrmions appear
frequently as clusters in large Skyrmions. We can see this by looking
at the charge 
$12$
sector. The \skfigref{12}{skfig:B12c6}-, \skfigref{12}{skfig:B12c282}-, \skfigref{12}{skfig:B12c73}-
and \skfigref{12}{skfig:B12c81}-Skyrmions look like two weakly
bound \skfigref{6}{skfig:B6c1}-Skyrmions;
the \skfigref{12}{skfig:B12c61}-Skyrmion like a $7+5$ cluster;
the \skfigref{12}{skfig:B12c60}-Skyrmion like a $8+4$ cluster;
the \skfigref{12}{skfig:B12c52}-, \skfigref{12}{skfig:B12c13}
and \skfigref{12}{skfig:B12c45}-Skyrmions like a $4+4+4$ cluster and
the \skfigref{12}{skfig:B12c15}-Skyrmion like a $3+4+5$ cluster. Some
of these solutions are shown in fig.~\ref{fig:12_clusters}. Often the
Skyrmions are attached symmetrically, so that some of their shared
symmetry group remains intact.

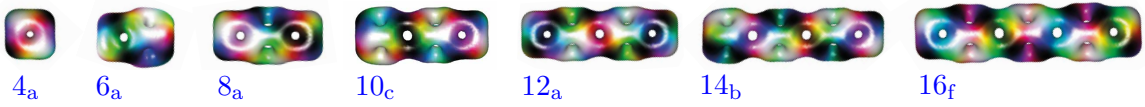
\begin{figure}[!h]
\centering
\begin{tikzpicture}
\node[inner sep=0] (image) at (0,0) {\includegraphics[width=1.0\textwidth]{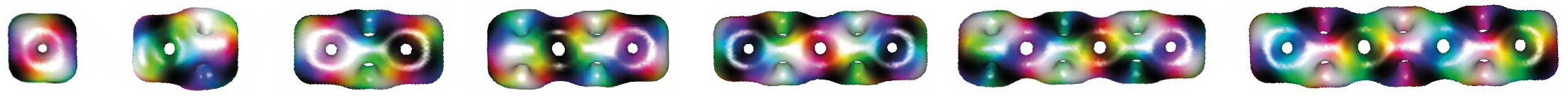}};
\node at (-7.2,-0.7) {\skfigref{4}{skfig:B4c2}};
\node at (-6.1,-0.7) {\skfigref{6}{skfig:B6c1}};
\node at (-4.5,-0.7) {\skfigref{8}{skfig:B8c1}};
\node at (-2.6,-0.7) {\skfigref{10}{skfig:B10c23}};
\node at (-0.4,-0.7) {\skfigref{12}{skfig:B12c51}};
\node at (1.95,-0.7) {\skfigref{14}{skfig:B14c47}};
\node at (4.8,-0.7) {\skfigref{16}{skfig:B16c122}};
\end{tikzpicture}
\caption{Straight chain Skyrmions from $B=4$ to $B=16$.}
\label{fig:straightchains}
\end{figure}

Many solutions are finite slices of an infinite (or large)
solution. The most studied infinite solution is the
chain \cite{Harland:2008eu}. There are several ways to interpret this:
as a chain of cubic \skfigref{4}{skfig:B4c2}-Skyrmions with
alternating orientations or as toroidal 2-Skyrmions stacked on their
axis of continuous
symmetry. The \skfigref{4}{skfig:B4c2}-, \skfigref{6}{skfig:B6c1}-, \skfigref{8}{skfig:B8c1}-, \skfigref{10}{skfig:B10c23}-, \skfigref{12}{skfig:B12c51}-, \skfigref{14}{skfig:B14c47}-
and \skfigref{16}{skfig:B16c122}-Skyrmions, seen in
fig.~\ref{fig:straightchains}, can be thought of as slices from this
chain. These solutions all have low energy and are the global
minimizers for $B=4,6$ and $12$. The energy of the 8-chain is
degenerate with the $D_{6d}$
symmetric \skfigref{8}{skfig:B8c3}-Skyrmion. For $B>12$, other
solutions have lower energy meaning that the chains become local,
rather than global, minimizers.

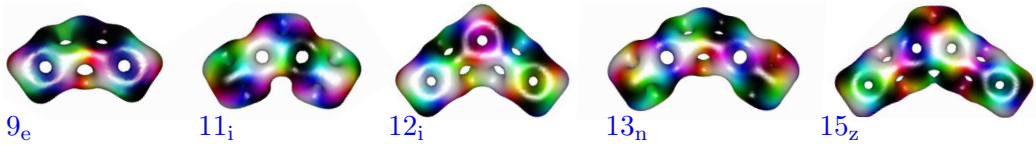
\begin{figure}[h!]
\centering
\begin{tikzpicture}
\node[inner sep=0] (image) at (0,0) {\includegraphics[width=0.9\textwidth]{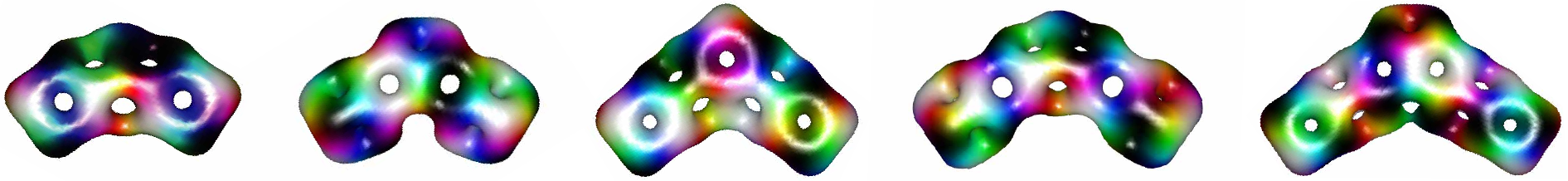}};
\node at (-6.6,-0.9) {\skfigref{9}{skfig:B9c13}};
\node at (-4,-0.9) {\skfigref{11}{skfig:B11c40}};
\node at (-1.5,-0.9) {\skfigref{12}{skfig:B12c45}};
\node at (1.4,-0.9) {\skfigref{13}{skfig:B13c42}};
\node at (4.2,-0.9) {\skfigref{15}{skfig:B15c31}};
\end{tikzpicture}
\caption{Hinge Skyrmions.}
\label{fig:hinges}
\end{figure}

The infinite chain can bend: a Skyrmion can be inserted
between two cubes, acting as a hinge. Different hinges give different
angles between the incoming and outgoing Skyrmions. We have found bent
chains with hinges given by
the \skfigref{1}{skfig:B1c1}-, \skfigref{3}{skfig:B3c1}-, \skfigref{4}{skfig:B4c2}-, \skfigref{5}{skfig:B5c1}-
and \skfigref{7}{skfig:B7c1}-Skyrmions, which can be seen in
fig.~\ref{fig:hinges}. These are
the \skfigref{9}{skfig:B9c13}-, \skfigref{11}{skfig:B11c40}-, \skfigref{12}{skfig:B12c45}-, \skfigref{13}{skfig:B13c42}-,
and \skfigref{15}{skfig:B15c31}-Skyrmions. Using these hinges, one
could construct a long Skyrmion with complex geometry. A simple
starting point would be a ring solution. Each hinge-type will generate
an optimal ring with a different size and baryon number. 

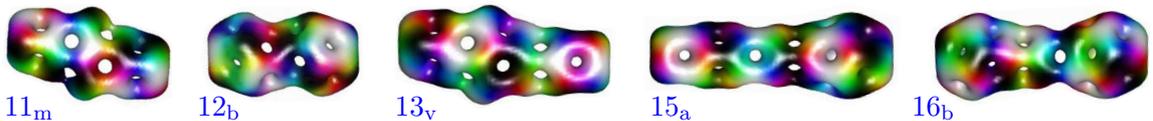
\begin{figure}[h!]
\centering
\begin{tikzpicture}
\node[inner sep=0] (image) at (0,0) {\includegraphics[width=1.0\textwidth]{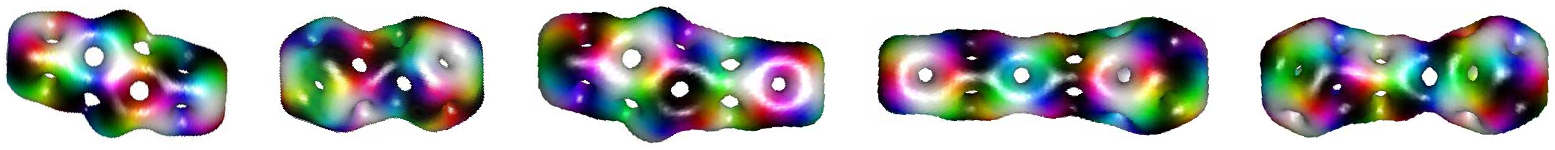}};
\node at (-7.2,-0.7) {\skfigref{11}{skfig:B11c22}};
\node at (-4.7,-0.7) {\skfigref{12}{skfig:B12c44}};
\node at (-2.1,-0.7) {\skfigref{13}{skfig:B13c76}};
\node at (1.25,-0.7) {\skfigref{15}{skfig:B15c6}};
\node at (4.7,-0.7) {\skfigref{16}{skfig:B16c72}};
\end{tikzpicture}
\caption{Chain Skyrmions made from 2- and 3-tori.}
\label{fig:2323}
\end{figure}

A new type of chain appears, and has low energy, for $B\geq11$. This
is constructed from a combination of $2$-tori and $3$-tori stacked
next to one another. The
dodecahedral \skfigref{7}{skfig:B7c1}-Skyrmion can also be interpreted
as a $2+3+2$ stack and configurations which repeat this pattern have
low energy. At $B=12$ the competing cubic chain and $2+3+2+3+2$
chain have very similar energies; while at $B=14$ the cubic
chain \skfigref{14}{skfig:B14c47}-Skyrmion has higher energy than the
new \skfigref{14}{skfig:B14c46}-Skyrmion, which takes the form of a
$2+3+2+2+3+2$ stack (or can be interpreted as
two \skfigref{7}{skfig:B7c1}-Skyrmions).
This suggests that the infinite $2+3$ chain will have lower energy
than the cubic
chain, which is interesting since the 3-torus is by itself unstable.
The \skfigref{11}{skfig:B11c22}-, \skfigref{12}{skfig:B12c90}-, \skfigref{13}{skfig:B13c76}-, \skfigref{15}{skfig:B15c6}-
and \skfigref{16}{skfig:B16c72}-Skyrmions are displayed in
fig.~\ref{fig:2323}. These chains are always capped by 2-tori,
suggesting that capping with a 3-torus is energetically
unfavorable. There is a particularly nice infinite series of Skyrmions 
with baryon number $7+5N$, which look like alternating $2$- and
$3$-tori capped by 2-tori. The \skfigref{7}{skfig:B7c1}-
and \skfigref{12}{skfig:B12c45}-Skyrmions take this form. The next in
the series is at $B=17$, beyond the scope of this paper.  

\begin{figure}[h!]
\centering
\begin{tikzpicture}
\node[inner sep=0] (image) at (0,0) {\includegraphics[width=0.7\textwidth]{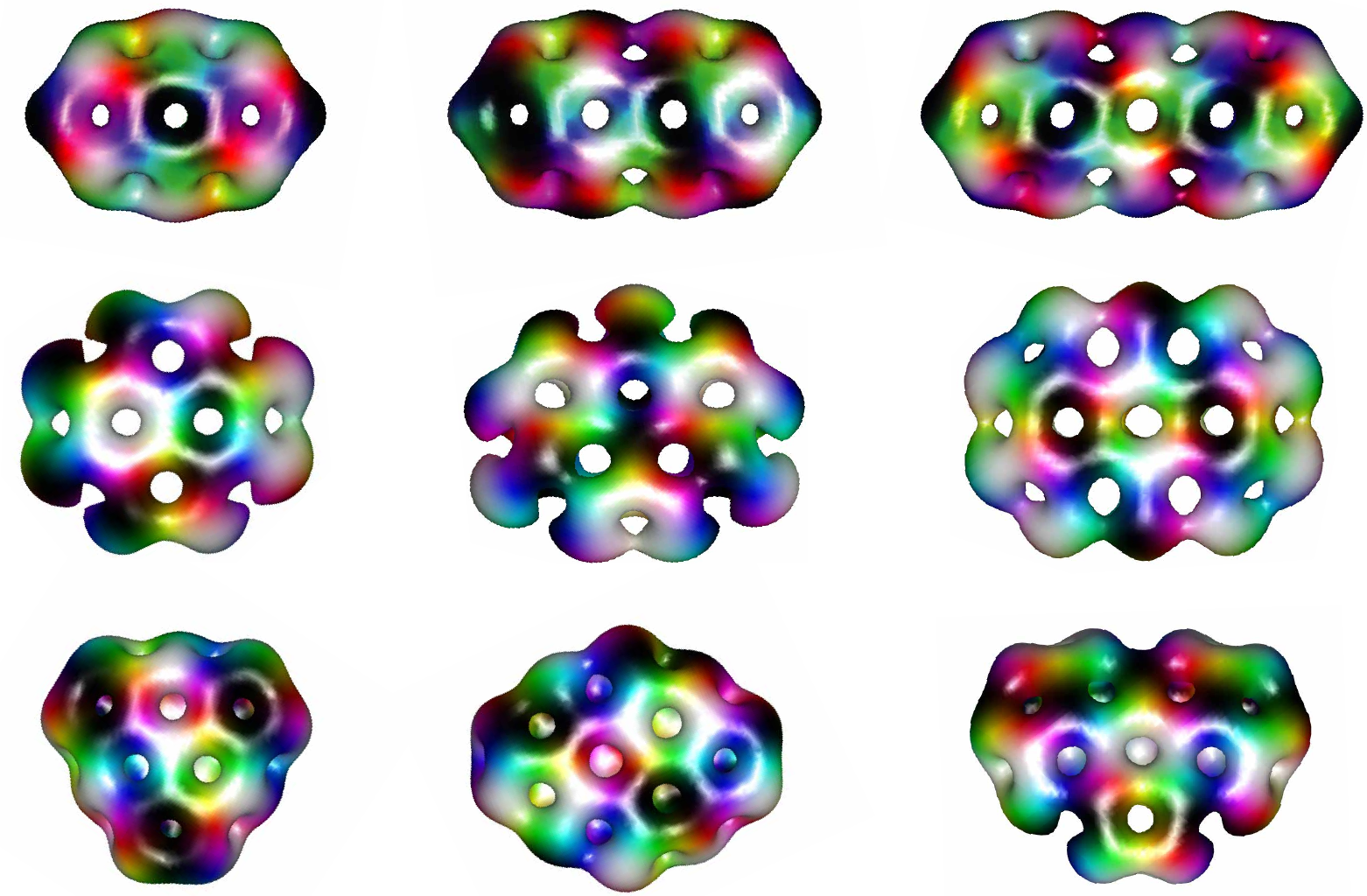}};
\node at (-4.9,1.8) {\skfigref{10}{skfig:B10c1}};
\node at (-1.7,1.8) {\skfigref{13}{skfig:B13c26}};
\node at (2,1.8) {\skfigref{16}{skfig:B16c4}};
\node at (-4.9,-0.8) {\skfigref{12}{skfig:B12c64}};
\node at (-1.7,-0.8) {\skfigref{14}{skfig:B14c51}};
\node at (2.3,-0.8) {\skfigref{16}{skfig:B16c110}};
\node at (-4.8,-3.2) {\skfigref{12}{skfig:B12c29}};
\node at (-1.6,-3.2) {\skfigref{14}{skfig:B14c27}};
\node at (2.4,-3.2) {\skfigref{15}{skfig:B15c71}};
\end{tikzpicture}
\caption{Graphene Skyrmions.}
\label{fig:graphenes}
\end{figure}
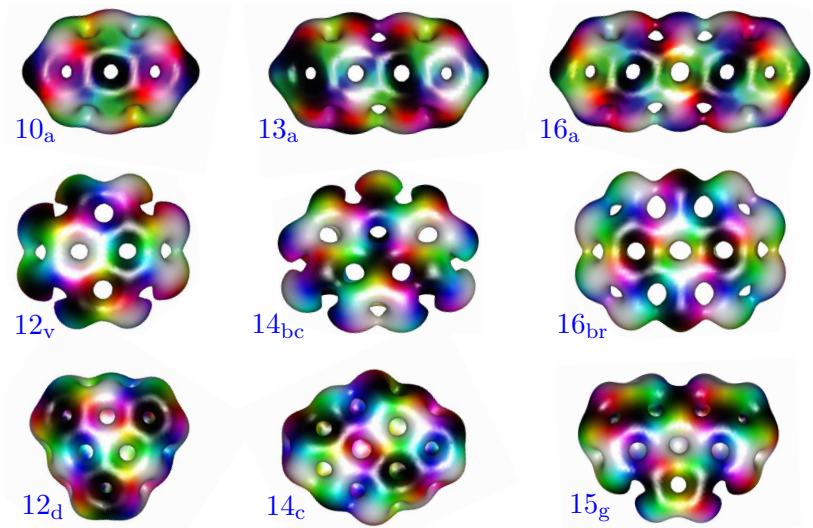

There is a well-known infinite flat lattice
solution \cite{Battye:1997wf}. Its energy is concentrated on a plane,
focused on the edges of tessellated hexagons with holes in the hexagon
centers. We call this solution graphene, due to its likeness with this
structure. Many large-$B$ Skyrmions look like two finite layers of
graphene stacked atop one another. Some examples are shown in
fig.~\ref{fig:graphenes}. The stacking can be done with the layers in
phase, as is seen for the top two rows of the figure, or out of phase
so that the holes of one layer match the vertices of the next layer,
as is seen in the bottom row. We call these two options symmetric and
out-of-phase graphene, respectively.  For some configurations the two
layers of graphene are not, or cannot, be stacked symmetrically. These
solutions often have no symmetry at all. The energy of the graphene
depends on the edge of the solution. The configurations in the top row
of fig.~\ref{fig:graphenes} seem to have a thicker edge than those in
the middle row. The edge of the graphene depends on how the finite
slice has been cut from the infinite solution. We posit that the thick
edge appears when the cut is made through the matter between two
holes. If a cut is made 
through a hole, single Skyrmions can appear. This can be seen in
the  \skfigref{14}{skfig:B14c51}-Skyrmion: at the top of this
solution, two cuts have taken place through two holes, leaving a
single Skyrmion jutting out. The next cuts are through the matter
between holes; leaving a smoother edge. The top row configurations are
all global minimizers for that baryon number, suggesting that
configurations with a thick edge will generically have lower
energy. They also appear to be the first three configurations in an
infinite sequence with baryon number $10+3N$. There are many more
examples of symmetric, out-of-phase and asymmetric graphene in the
table of Skyrmions (see app.~\ref{app:sols}). 

Although the one-layer graphene solution has been studied in detail,
our finite $B$ solutions suggest that this structure rarely appears by
itself in Skyrmions. In contrast, the two-layer solution appears
regularly. Hence, it seems worthwhile to do a study of infinite
multilayered graphene. An initial study was done for $m=0$
in ref.~\cite{SilvaLobo:2009qw}. It is also worth noting that even small
solutions can be interpreted as graphene. The cubic 4-Skyrmion is a
tightly folded slice of graphene with 6 holes. One could think of the
graphene as an infinite piece of paper, and could construct many
Skyrmions by cutting out nets and folding them together into
solutions. 

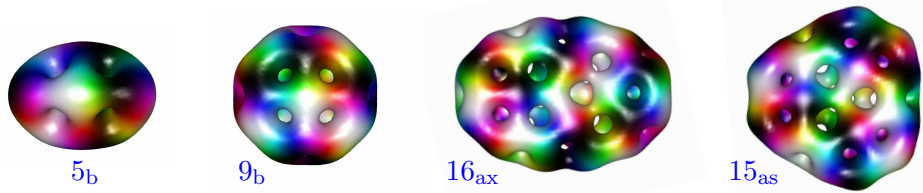
\begin{figure}[h!]
\centering
\begin{tikzpicture}
\node[inner sep=0] (image) at (0,0) {\includegraphics[width=0.8\textwidth]{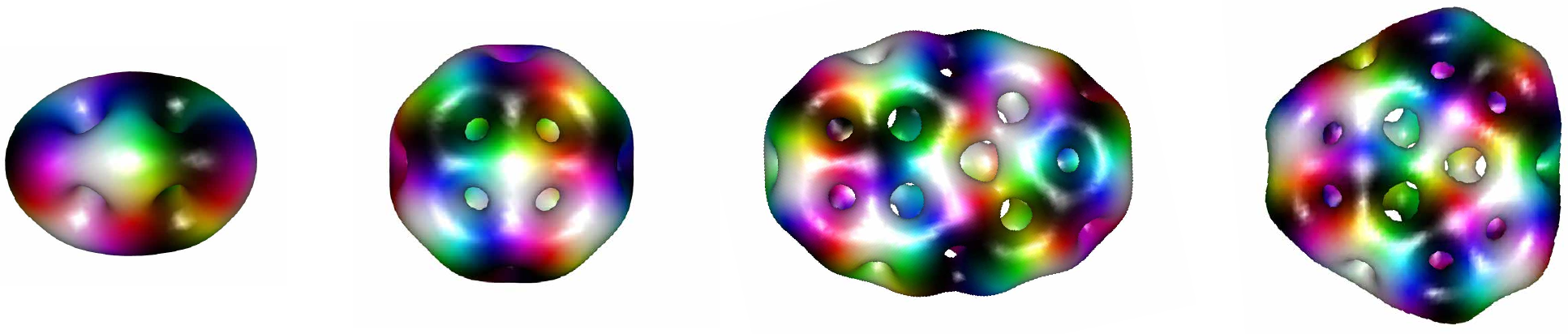}};
\node at (-5,-1) {\skfigref{5}{skfig:B5c5}};
\node at (-2.8,-1) {\skfigref{9}{skfig:B9c2}};
\node at (0.1,-1) {\skfigref{16}{skfig:B16c68}};
\node at (3.8,-1) {\skfigref{15}{skfig:B15c51}};
\end{tikzpicture}
\caption{Layered Skyrmions with tetravalent vertices.}
\label{fig:tetravalent}
\end{figure}

The hexagonal lattice has been known for a long time. Battye and
Sutcliffe noted that the solution looks like an infinite array of
trivalent vertices. Using this observation they proposed the GEM rules \cite{Battye:1997qq}
for Skyrmion energetics, which state that low energy Skyrmions will
have an energy density consisting of these trivalent vertices. The
global minimizers, which include some configurations from
fig.~\ref{fig:2323} and some from the top row of
fig.~\ref{fig:graphenes}, all obey this rule. However, we have also
found solutions with tetravalent vertices. This is first seen for the
new \skfigref{5}{skfig:B5c5}-Skyrmion, and the structure can be seen
in other solutions such as
the \skfigref{9}{skfig:B9c2}-, \skfigref{15}{skfig:B15c51}-
and \skfigref{16}{skfig:B16c68} Skyrmions. These solutions can be seen
in fig.~\ref{fig:tetravalent}. Apart from
the \skfigref{5}{skfig:B5c5}-Skyrmion, these solutions all have
relatively high energy, agreeing with the GEM rule energetics. 

\begin{figure}[h!]
\centering
\begin{tikzpicture}
\node[inner sep=0] (image) at (0,0) {\includegraphics[width=0.8\textwidth]{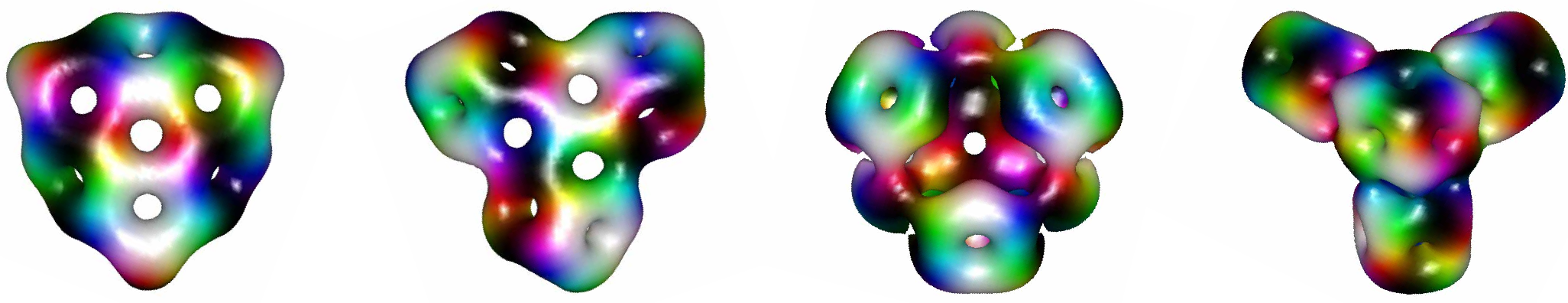}};
\node at (-6,-1) {\skfigref{11}{skfig:B11c39}};
\node at (-2.7,-1) {\skfigref{13}{skfig:B13c63}};
\node at (0.4,-1) {\skfigref{15}{skfig:B15c134}};
\node at (3.8,-1) {\skfigref{16}{skfig:B16c10}};
\end{tikzpicture}
\caption{Skyrmions with at least $C_3$ symmetry.}
\label{fig:C3}
\end{figure}
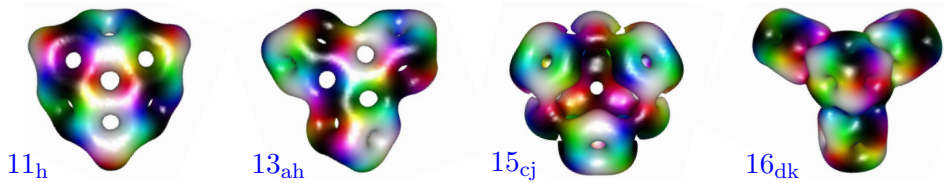

High-symmetry solutions have always motivated studies of classical and
quantized Skyrmions. We do not find many new high-symmetry
solutions. Some examples are the
$D_{5d}$ \skfigref{12}{skfig:B12c44}- and \skfigref{14}{skfig:B14c46}-Skyrmions,
the $D_{2h}$ \skfigref{15}{skfig:B15c109}-Skyrmion and the
$C_{3v}$ \skfigref{16}{skfig:B16c10}-Skyrmion. The last is one of many
solutions with at least $C_3$ symmetry. Some more are displayed in
fig.~\ref{fig:C3}. Here we see that $C_3$ symmetry can appear in
Skyrmions which look like graphene, cubes sharing a single Skyrmion, a
flat arrangement containing tori or simply cubes. No single type of
solution has a monopoly on any one symmetry group. 

\begin{figure}[h!]
\centering
\begin{tikzpicture}
\node[inner sep=0] (image) at (0,0) {\includegraphics[width=0.7\textwidth]{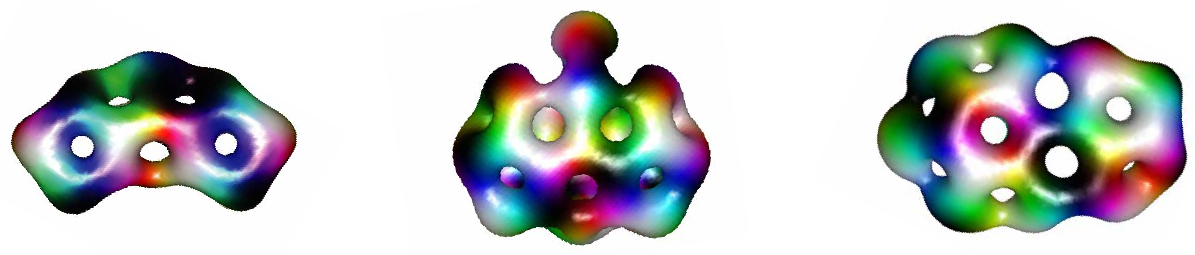}};
\node at (-4.9,-1) {\skfigref{9}{skfig:B9c13}};
\node at (-1.2,-1) {\skfigref{11}{skfig:B11c36}};
\node at (2.8,-1) {\skfigref{12}{skfig:B12c59}};
\end{tikzpicture}
\caption{Solutions whose constituent clusters aren't solutions themselves.}
\label{fig:stable}
\end{figure}
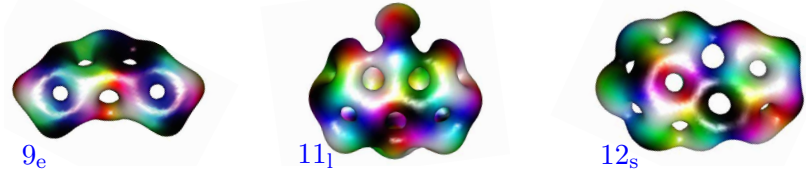

Occasionally the presence of one Skyrmion can stabilize another,
previously unstable configuration. An example of this is the untwisted
$B=8$ solution. This is a well-known saddle point of the Skyrme model,
and it looks like two cubes next to each other in the same
orientation. A small mode analysis shows that the solution is unstable
to one of the cubes twisting on the axis joining the
pair \cite{Gudnason:2018bju}. However,
the \skfigref{9}{skfig:B9c13}-Skyrmion looks like two cubes in the
same orientation, but fused by
a \skfigref{1}{skfig:B1c1}-Skyrmion. The \skfigref{1}{skfig:B1c1}-Skyrmion
somehow stabilizes the previously unstable configuration. Another
example is the \skfigref{11}{skfig:B11c36}-Skyrmion, which looks like
a slice of out-of-phase graphene with
a \skfigref{1}{skfig:B1c1}-Skyrmion attached. There is no
$10$-Skyrmion which looks like out-of-phase graphene, meaning that the
$1$-Skyrmion has stabilized the configuration. This out-of-phase structure
with baryon number 10 is also seen in
the \skfigref{14}{skfig:B14c39}, \skfigref{15}{skfig:B15c49}, \skfigref{16}{skfig:B16c19}
and \skfigref{16}{skfig:B16c120} solutions. Finally,
the  \skfigref{12}{skfig:B12c59}-Skyrmion looks like a(n unstable)
$D_{6h}$ 8-Skyrmion attached to another cluster. These examples can be
seen in fig.~\ref{fig:stable}. 

We have so far focused on special solutions of the Skyrme
model. However, most other solutions can be understood as some
combination or deformation of those we've discussed. For instance:
the \skfigref{12}{skfig:B12c32}-, \skfigref{13}{skfig:B13c3}-, \skfigref{14}{skfig:B14c157}-
and \skfigref{16}{skfig:B16c94}-Skyrmions are all almost-symmetric
graphene;
the \skfigref{12}{skfig:B12c7}-, \skfigref{14}{skfig:B14c17}-, \skfigref{15}{skfig:B15c49}-
and \skfigref{16}{skfig:B16c31}-Skyrmions look like small-$B$
Skyrmions attached to graphene;
the \skfigref{12}{skfig:B12c49}-, \skfigref{14}{skfig:B14c37}-, \skfigref{15}{skfig:B15c13}-
and \skfigref{16}{skfig:B16c115}-Skyrmions are deformed graphene with
no symmetry and
the \skfigref{11}{skfig:B11c222}-, \skfigref{12}{skfig:B12c37}-, \skfigref{14}{skfig:B14c118}-
and \skfigref{15}{skfig:B15c184}-Skyrmions are deformed chains. This
is only a small collection of examples. Using a combination of small
solutions, graphene, chains and hinges we can describe almost all of
the \numsols{} Skyrmions found. There can also be many different
versions of the same combination. For instance
the \skfigref{16}{skfig:B16c209}-, \skfigref{16}{skfig:B16c198}-, \skfigref{16}{skfig:B16c67}-, \skfigref{16}{skfig:B16c104}-, \skfigref{16}{skfig:B16c174}-Skyrmions
all look like a \skfigref{6}{skfig:B6c1}-Skyrmion attached to
symmetric graphene and have near-degenerate energy. However, they all
look different as there are many ways to combine the two objects.

It is worth noting that many solutions belong to several of these
categories and can be interpreted in different ways. For instance
the \skfigref{12}{skfig:B12c6}-Skyrmion looks like
two \skfigref{6}{skfig:B6c1}-Skyrmions attached
symmetrically \emph{and} like a slice of symmetric graphene. This is
equally true of small Skyrmions: the
cubic \skfigref{4}{skfig:B4c2}-Skyrmion looks like two 2-tori, the
dodecahedral \skfigref{7}{skfig:B7c1}-Skyrmion like a 3-torus
sandwiched between two 2-tori and so on. Our descriptions in this
Section and in app.~\ref{app:sols} are suggestions of how to
interpret the solutions. Depending on the application, certain
interpretations may be more or less appropriate.

\subsection{Soft modes}\label{sec:softmodes}

Our numerical scheme is quite accurate and this makes it easy for the
algorithm explained in sec.~\ref{sec:elimination} to find and
eliminate doublets, i.e.~solutions found previously.
The main exception to this is caused by a phenomenon that we call
``soft modes'', which are almost-zero modes in that they have a very
small excitation energy.
Alternatively, they can be thought of as a path between two or more
solutions in configuration space with an extremely shallow potential
barrier. 
A good example of this is the \skfigref{12}{skfig:B12c61}-Skyrmion,
which contains a soft mode as it takes very little energy to rotate
the \skfigref{5}{skfig:B5c1}-Skyrmion relative to 
the \skfigref{7}{skfig:B7c1}-Skyrmion.
Another example in the same topological sector are
the \skfigref{12}{skfig:B12c81}-
and \skfigref{12}{skfig:B12c73}-Skyrmions, for which the relative
rotation of the \skfigref{6}{skfig:B6c1}-Skyrmion is a very low-energy
mode.
There are many examples with high energies, to just mention a few,
the \skfigref{13}{skfig:B13c73}-, \skfigref{14}{skfig:B14c71}-, \skfigref{14}{skfig:B14c83}-, \skfigref{14}{skfig:B14c79}-, \skfigref{14}{skfig:B14c134}-, \skfigref{15}{skfig:B15c75}-, \skfigref{14}{skfig:B15c94}-, \skfigref{15}{skfig:B15c72}-, \skfigref{15}{skfig:B15c22}-, \skfigref{15}{skfig:B15c64}-, \skfigref{15}{skfig:B15c133}-, \skfigref{16}{skfig:B16c52}-, \skfigref{16}{skfig:B16c2}-Skyrmions
contain soft modes. We have generally deleted 
solutions which are related by soft modes unless there is a compelling
reason, such as enhanced symmetries, to keep them.

\subsection{Comparison to past results}

Our results provide the most comprehensive search for solutions of the
Skyrme model to date. As such, we will now compare our results to past
studies, predictions and approximations.

The Rational Map Approximation (RMA) describes Skyrmions as shell-like
configurations, which contain a center filled with antivacuum
($U=-\mathbf{1}_2$). As the pion mass $m$ increases so does the energy
cost of this center. As such, we expect that prediction from the RMA
do not compare well to the results here, where $m=1$. The RMA
successfully describes the fullerene solutions for
$B=1$-$8$ \cite{Houghton:1997kg}. For $B=9$-$16$, Battye and Sutcliffe
calculated the optimal RMA
Skyrmions, and relaxed them numerically \cite{Battye:2000se,Battye:2001qn}. Of these, we only find
the $C_{4d}$ symmetric \skfigref{9}{skfig:B9c2}-Skyrmion and the
$D_{3h}$ \skfigref{11}{skfig:B11c39}-Skyrmion. There is also a
higher-energy tetrahedral \skfigref{9}{skfig:B9c44}-Skyrmion. None are
the global energy minimizers. The other RMA configurations are not seen
and may be saddle points of the theory. We also do not see any of the
proposed Skyrmions from the double RMA \cite{Manton:2000kj}, such as
the tetrahedral $B=12$ and cubic $B=13$ configurations.

In ref.~\cite{Battye:2006tb}, the authors used squashed RMA
Skyrmions as initial conditions to find solutions of the Skyrme model
with $m=1$. They found flat solutions for $B=10$-$16$ which we would
call graphene. We find all these
configurations. The \skfigref{10}{skfig:B10c1}-Skyrmion is the global
minimizer while
the \skfigref{11}{skfig:B11c30}, \skfigref{12}{skfig:B12c29}, \skfigref{13}{skfig:B13c3}, \skfigref{14}{skfig:B14c27}, \skfigref{15}{skfig:B15c12}
and \skfigref{16}{skfig:B16c7} are the third, fourth, eleventh,
third, third and third lowest energy solutions in their respective
baryon sectors.

In the $\alpha$-particle model, $B=4N$ Skyrmions are constructed from
$N$ $B=4$
cubes \cite{Battye:2006na}.
In refs.~\cite{Feist:2012ps,Lau:2014baa,Halcrow:2016spb}
the authors construct highly symmetric $\alpha$-clusters for $B=12$
and $16$. We do reproduce the \skfigref{12}{skfig:B12c51} chain,
the \skfigref{12}{skfig:B12c52} triangle,
the \skfigref{16}{skfig:B16c122} chain and
the \skfigref{16}{skfig:B16c79} bent square. However, we find no
tetrahedral $B=16$ solutions, despite there being many ways to
construct such a state. The closest configuration to this is
the \skfigref{16}{skfig:B16c10}-Skyrmion which is a flattened tetrahedron
with $C_{3v}$ symmetry. The fact that we do not find a $B=16$ tetrahedron
does not mean it is not an important configuration: it could be a low
energy saddle point.

Finally, we do not find any of the previously discovered saddle points
of the Skyrme model. These include the $B=2$ hedgehog, $B=5$
$O_h$-symmetric solution, $B=8$ untwisted chain and $B=16$ flat
square. That we do not find these is a reassuring check on our
numerical scheme, which is designed to only find minimizers.

\subsection{Statistics}

\begin{figure}[!htp]
\begin{center}
\mbox{\subfloat[]{\includegraphics[width=0.49\linewidth]{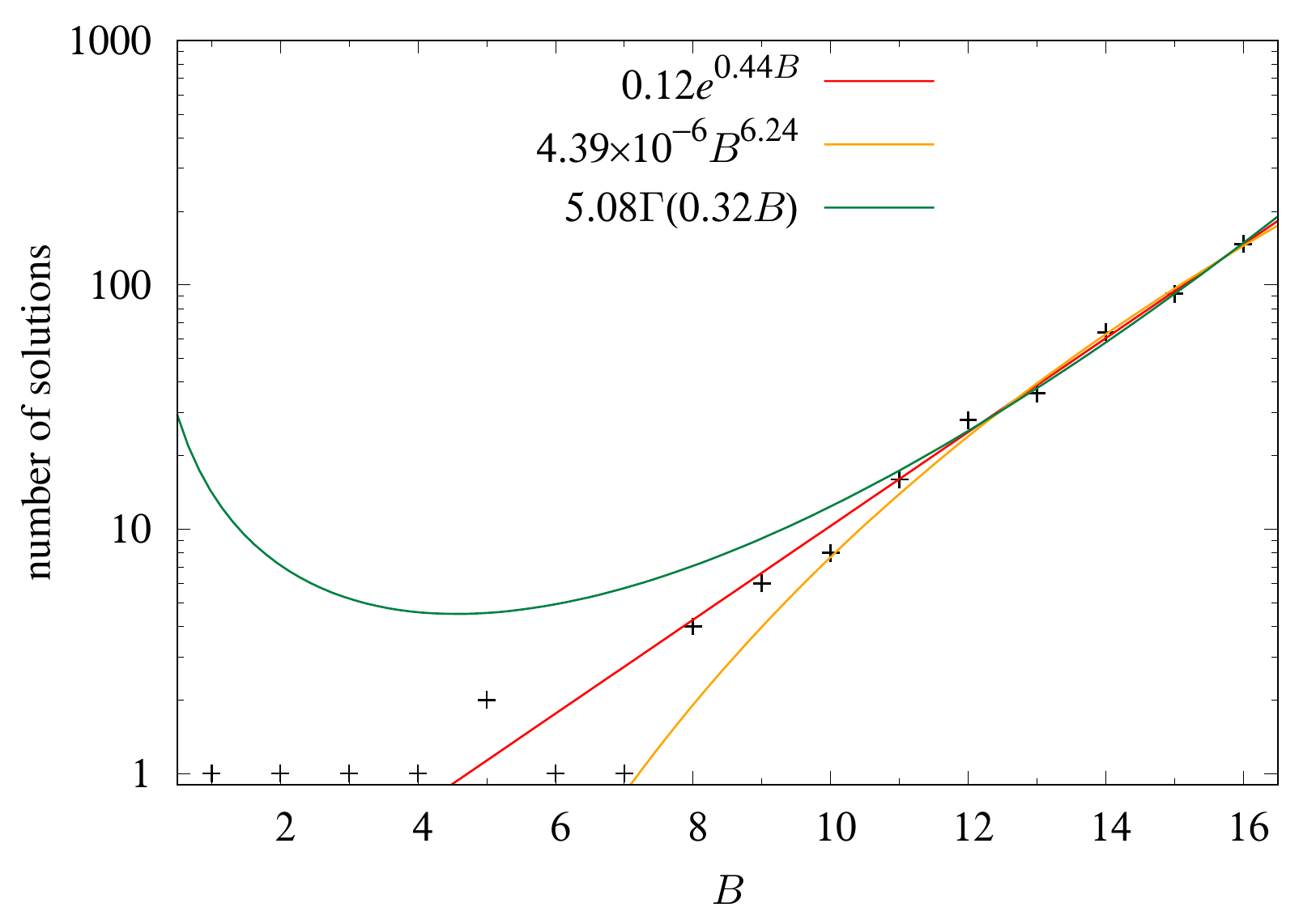}}
\subfloat[]{\includegraphics[width=0.49\linewidth]{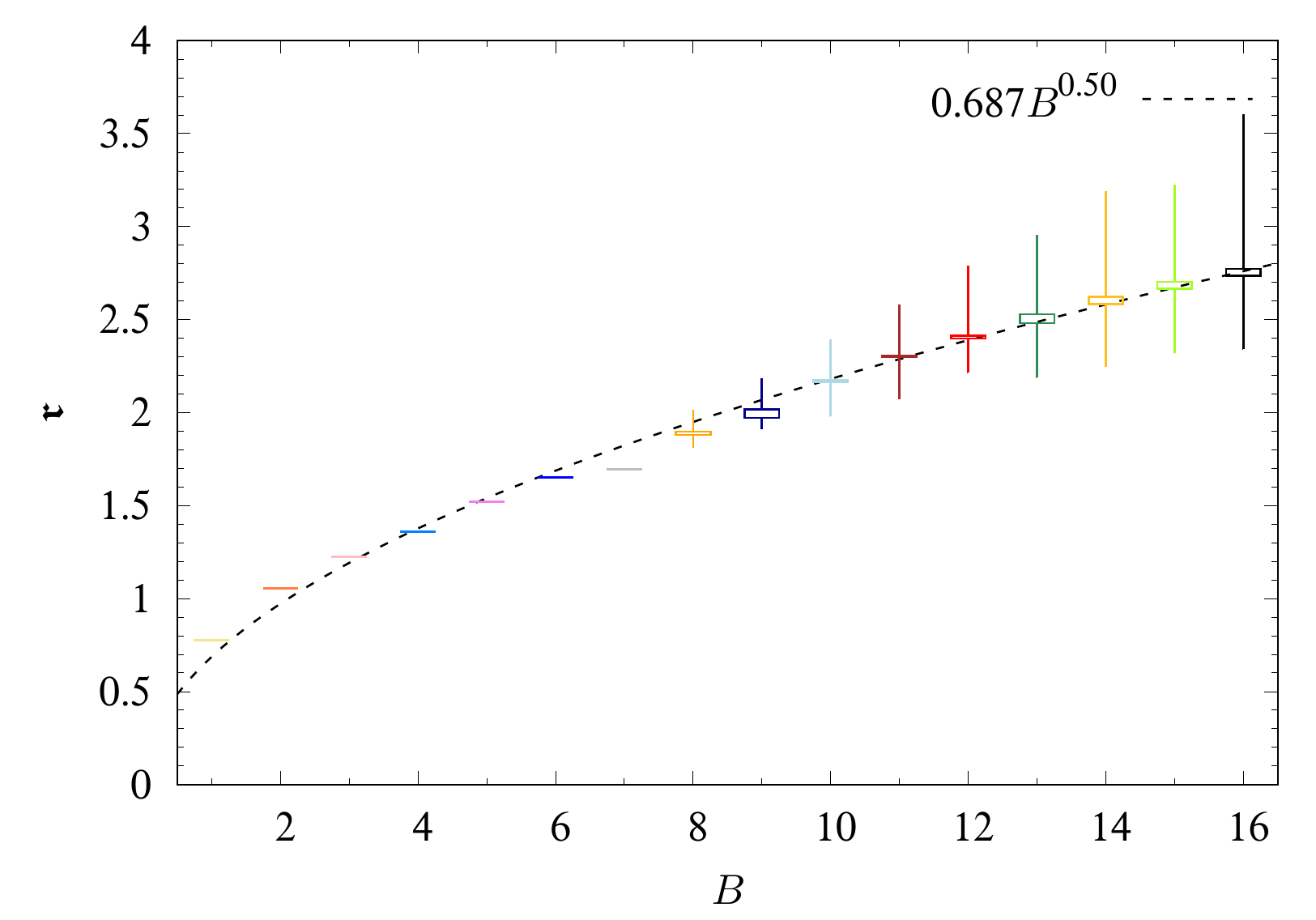}}}
\caption{(a) Number of solutions as a function of $B$ and (b) mean
(upper edge of box) and median (lower edge of box) of radii for
each $B$. The spread of the radius is shown with a solid line above
and below the box. }
\label{fig:stat_nr_sols}
\end{center}
\end{figure}

Having obtained \numsols{} solutions of the Skyrme model, it is
sensible to calculate some macro properties of the set of
solutions. First, we consider how many solutions there are
for a given baryon number $B$, presented in
fig.~\ref{fig:stat_nr_sols}(a).
With the data we have at this point, it is difficult to determine the
type of growth, but at least polynomial growth seems likely. Note
that the polynomial growth suggests that there would be $\sim 40,000$
solutions at $B=40$. Hence finding all energy minimizers for important
nuclei such as Calcium-40 would be prohibitively difficult.
We have also shown the average and median of the radii as well as the
spread of the radii for each $B$ in fig.~\ref{fig:stat_nr_sols}(b).
A fit shows that the mean radius of the Skyrmions is roughly
proportional to $\sqrt{B}$. Curiously, this is midway between the
behavior of normal materials ($R\sim M^{1/3}$) and BPS monopoles or
black holes ($R\sim M$) \cite{Bolognesi:2010xt}, where we have used
that $B\sim M$ with $M$ being the (baryon) mass.

\begin{figure}[!htp]
\begin{center}
\mbox{\subfloat[]{\includegraphics[width=0.49\linewidth]{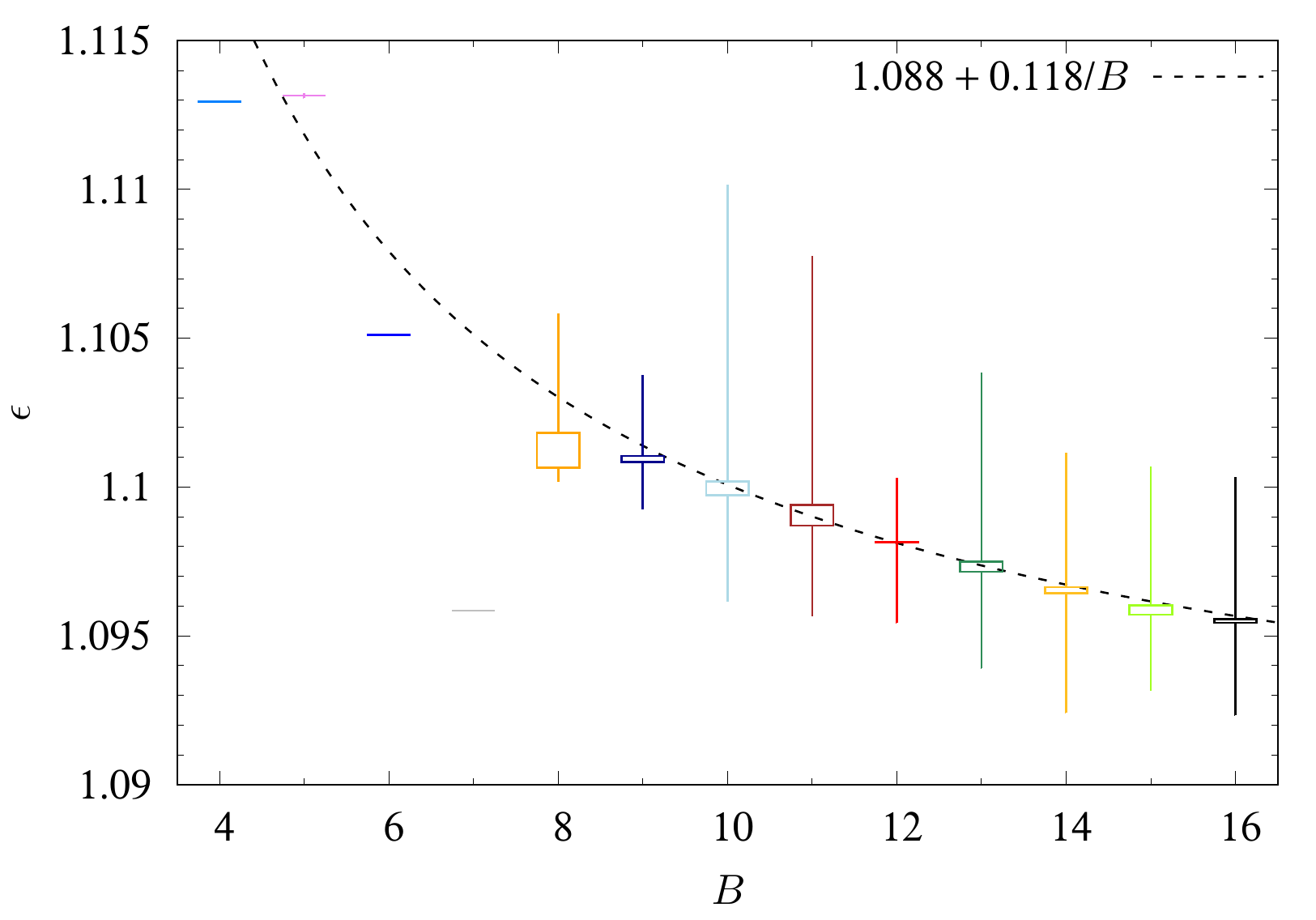}}
\subfloat[]{\includegraphics[width=0.49\linewidth]{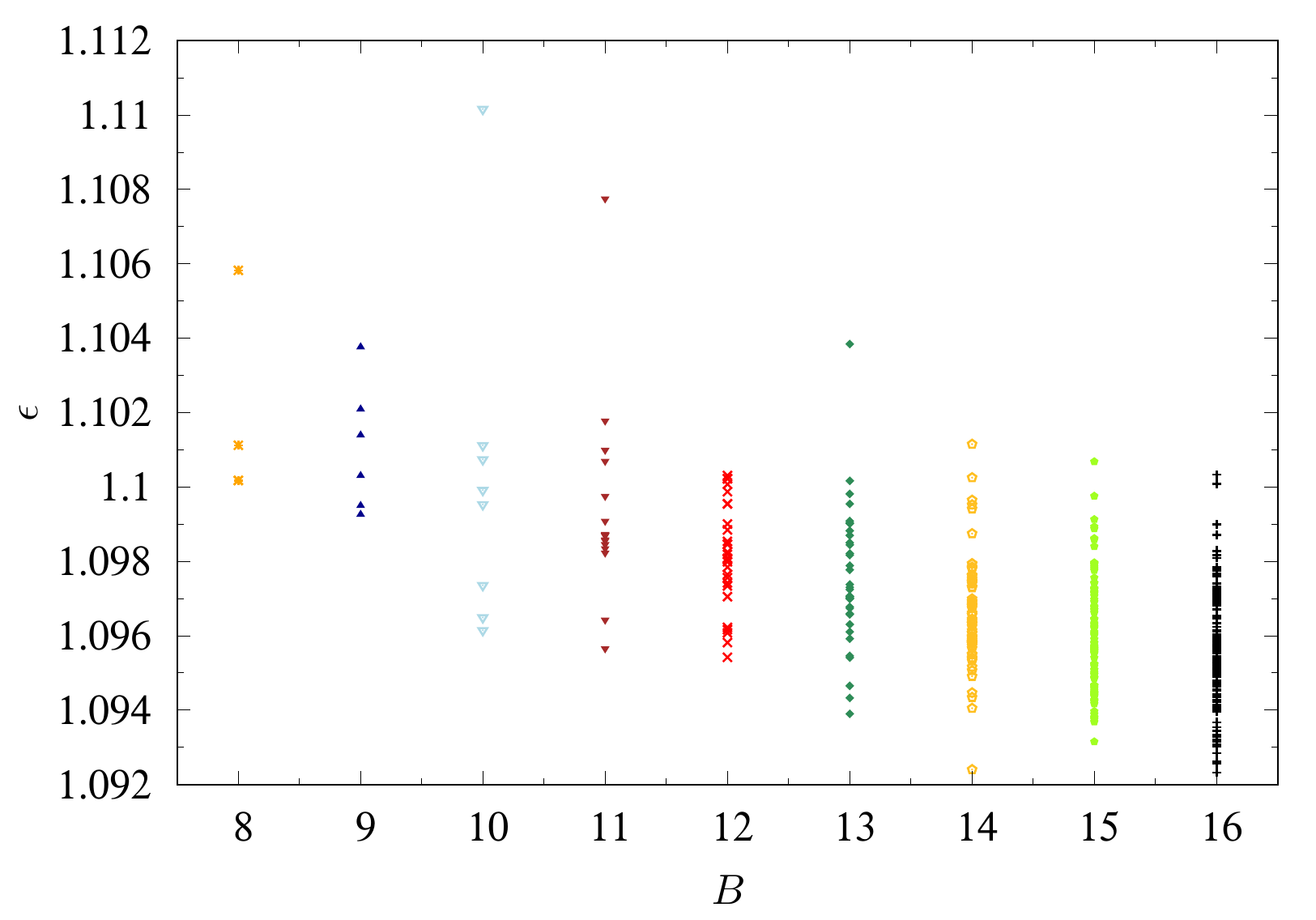}}}
\caption{Energy normalized by the energy bound and divided by the
baryon number $B$. That is $\epsilon$ of
eq.~\eqref{eq:epsilon_def}. (a) The mean and median as a box with
the lines representing the spread. (b) Scatter plot of the energies. }
\label{fig:stat_energy}
\end{center}
\end{figure}

We plot the normalized energies (normalized by the energy bound) per
baryon number $\epsilon$ (see eq.~\eqref{eq:epsilon_def}) in
fig.~\ref{fig:stat_energy}. This shows that the smallest $\epsilon$
generally decreases as $B$ increases, as expected. One exception is
the minimizing \skfigref{14}{skfig:B14c46}-Skyrmion which has very low
energy and a large energy gap. This naively matches the exceptionally
low binding energy of Nitrogen-14. We also see several anomalously
high energy solutions, for $B=10,11,13$. If such solutions exist for
these baryon numbers, they might also exist for other $B$. This
suggests that there could be solutions that our method has missed. We
do not claim to have found \emph{all} Skyrmions. Note, the energies
plotted in fig.~\ref{fig:stat_energy} are all classically stable
solutions; that is, no saddle points are included in our data.

\begin{figure}[!htp]
\begin{center}
\includegraphics[width=0.8\linewidth]{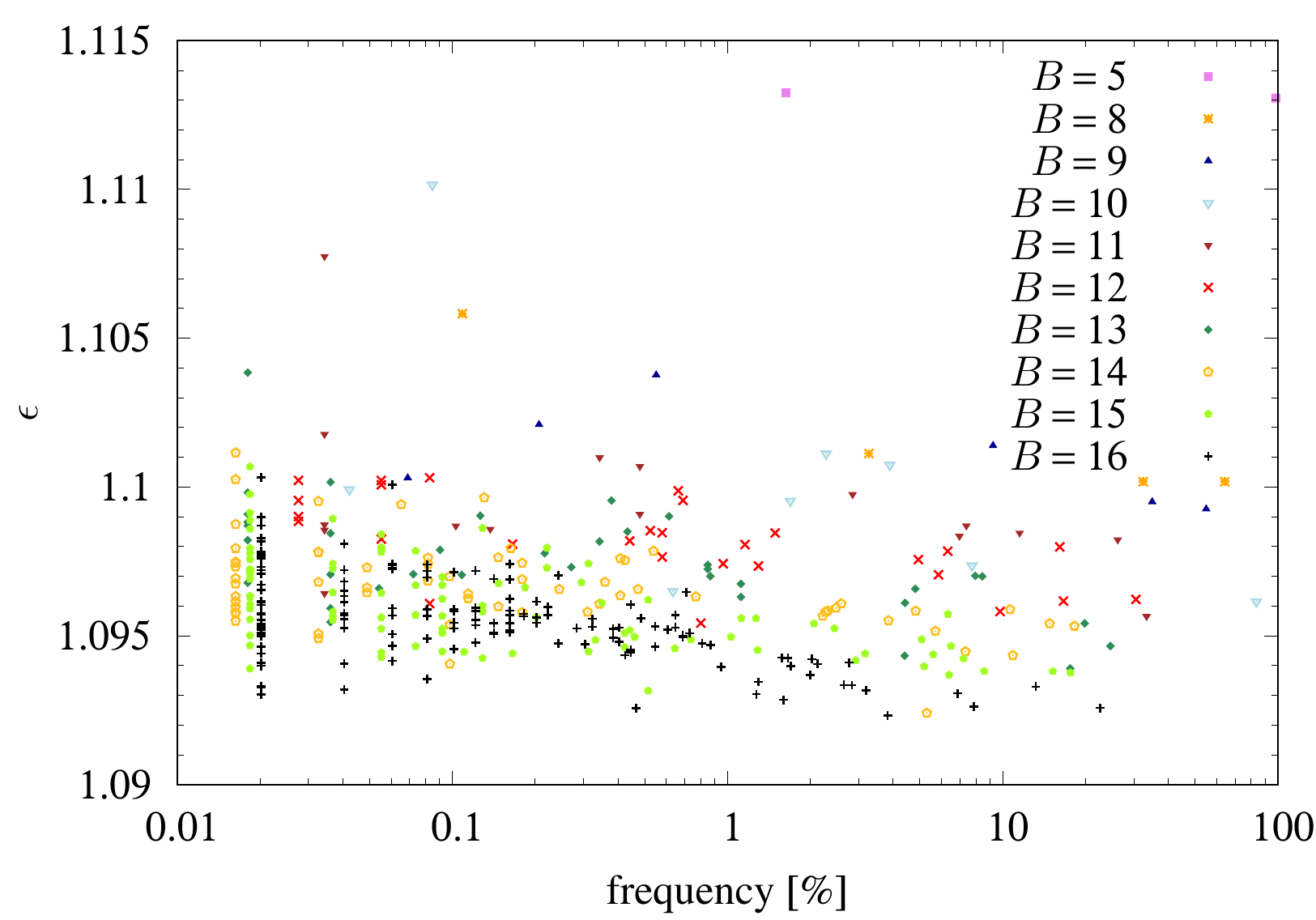}
\caption{Frequency of a particular solution being found by the
pseudo-random algorithm.}
\label{fig:stat_freq}
\end{center}
\end{figure}

The random initial conditions lead to a large number of solutions,
many of which are identical. Although our algorithm eliminates
repeated Skyrmions, we keep every log file and have 
tracked how many times the same solution has been found. This tells us
the frequency of a solution, which might indicate the
size of the basin of attraction of the solution. 
The result is shown in fig.~\ref{fig:stat_freq}.
For small $B$, the lowest energy
solution is the most common solution out of randomly generated initial
configurations; for large $B\gtrsim 12$, this is no longer the
case and the global minimizer of the energy functional is actually quite
rare. Equally, there are plenty of high energy solutions
which are quite common. These results may be showing the bias of our
initial configuration generator. For instance long chains are very
rare, and it is also very rare for a random initial configuration to
take this shape.
There are many solutions for $B\geq11$ that we only find
once. If we have found several solutions which are very rare, it is
logical to assume that there are even more still
undiscovered. Hence, we have likely not found all solutions for
$B\geq 11$.

\begin{figure}[!htp]
\begin{center}
\includegraphics[width=0.8\linewidth]{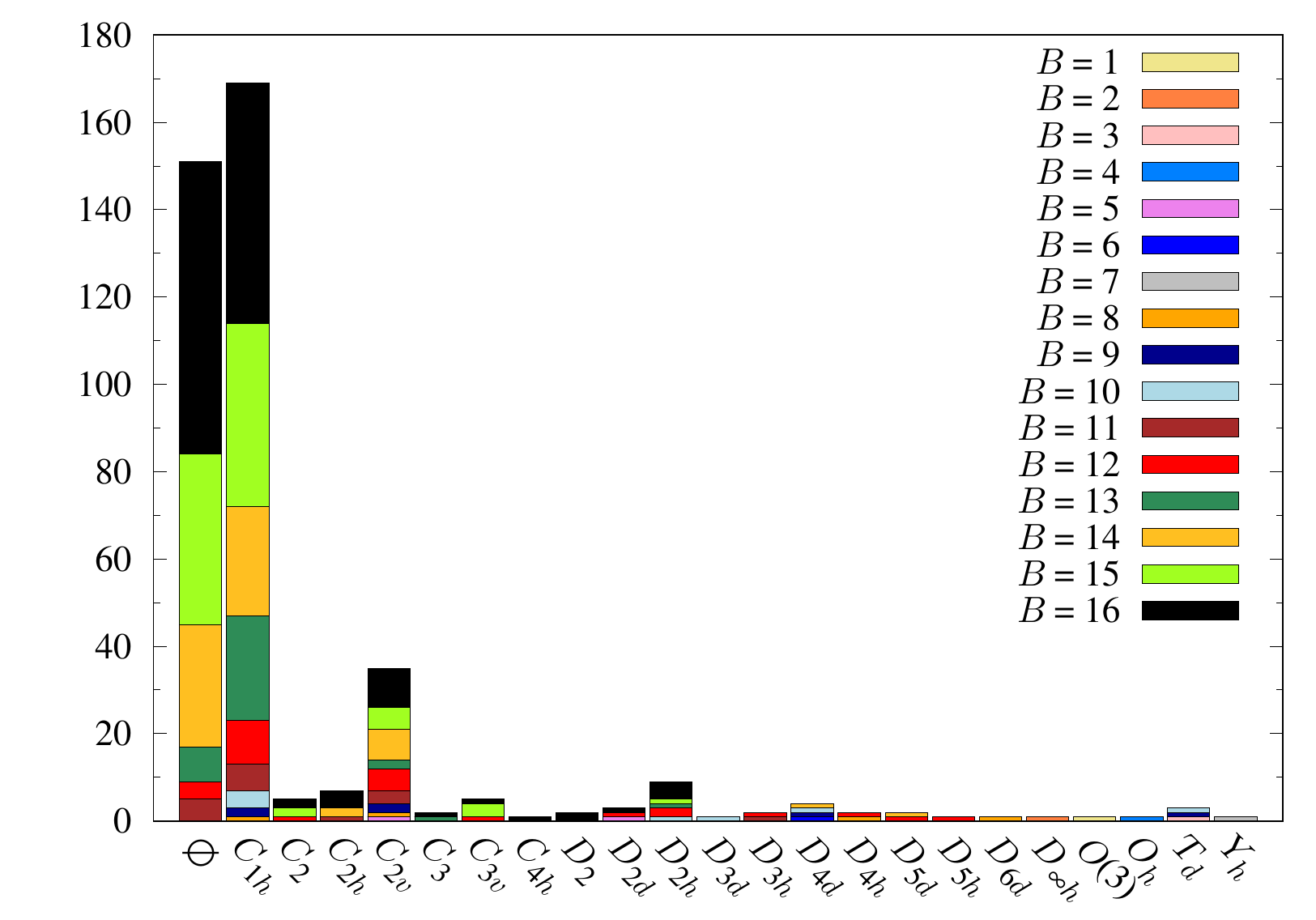}
\caption{Number of solutions possessing certain symmetries.}
\label{fig:stat_sym}
\end{center}
\end{figure}

In fig.~\ref{fig:stat_sym}, we plot the number of solutions with a 
given symmetry. The colors (globally defined in this section)
represent the solution of a given baryon number $B$. We see that the
most common symmetry groups are the empty set and $C_{1h}$ (one
reflection). The next most common is $C_{2v}$ (two reflections). Hence
a slim majority of Skyrmions do have a reflection symmetry. The fact
that so many solutions have no symmetry goes against the common lore
that Skyrmions have high symmetry. Many of these asymmetric solutions
contain sub-clusters which do have approximate symmetry. Perhaps
these symmetries should be taken into account in a full
quantization. We note that in the $\alpha$-particle model of 
nuclear physics, the individual sub-clusters are quantized first and
then interactions and dynamics are considered.

\begin{figure}[!htp]
\begin{center}
\mbox{\subfloat[]{\includegraphics[width=0.49\linewidth]{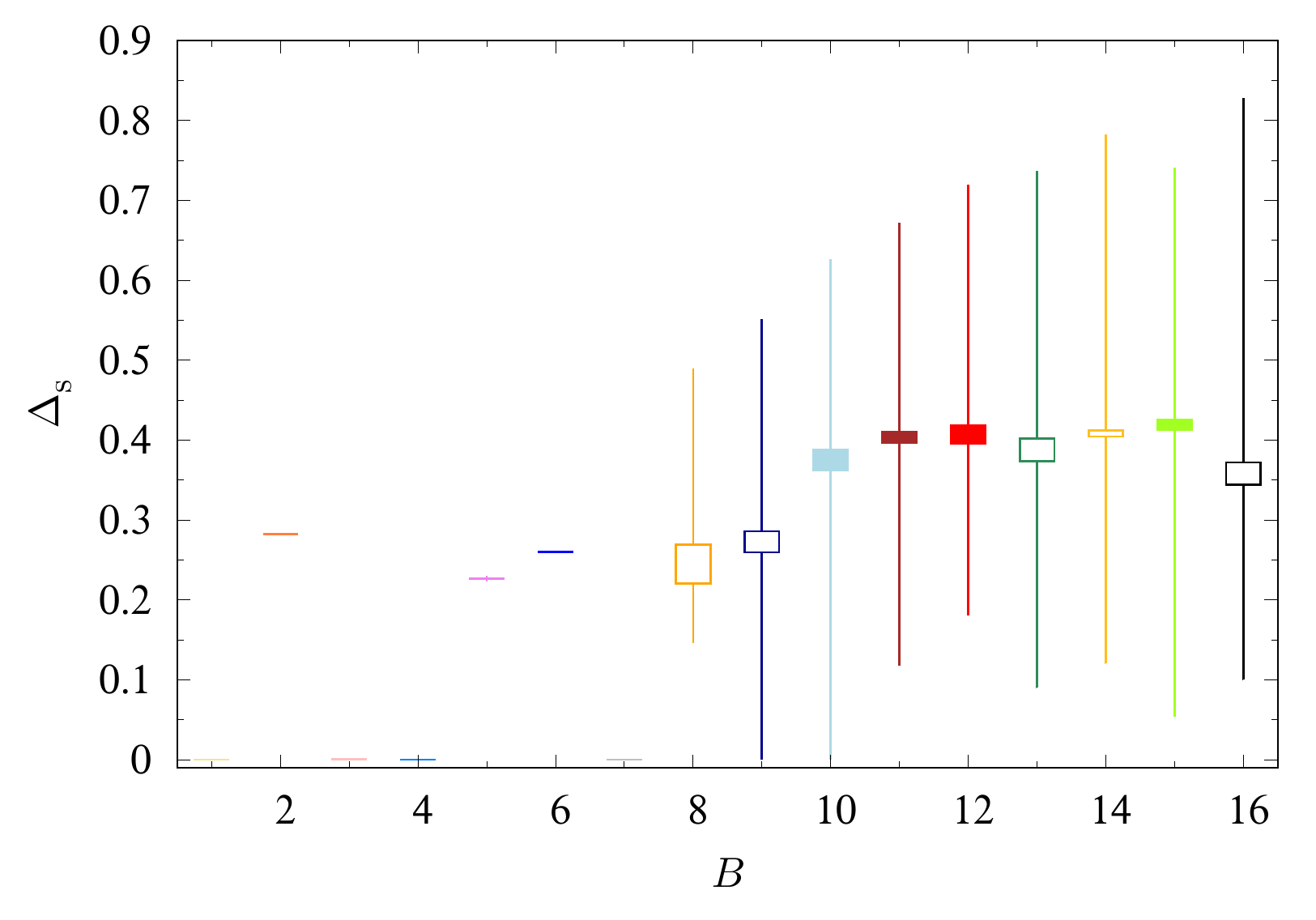}}
\subfloat[]{\includegraphics[width=0.49\linewidth]{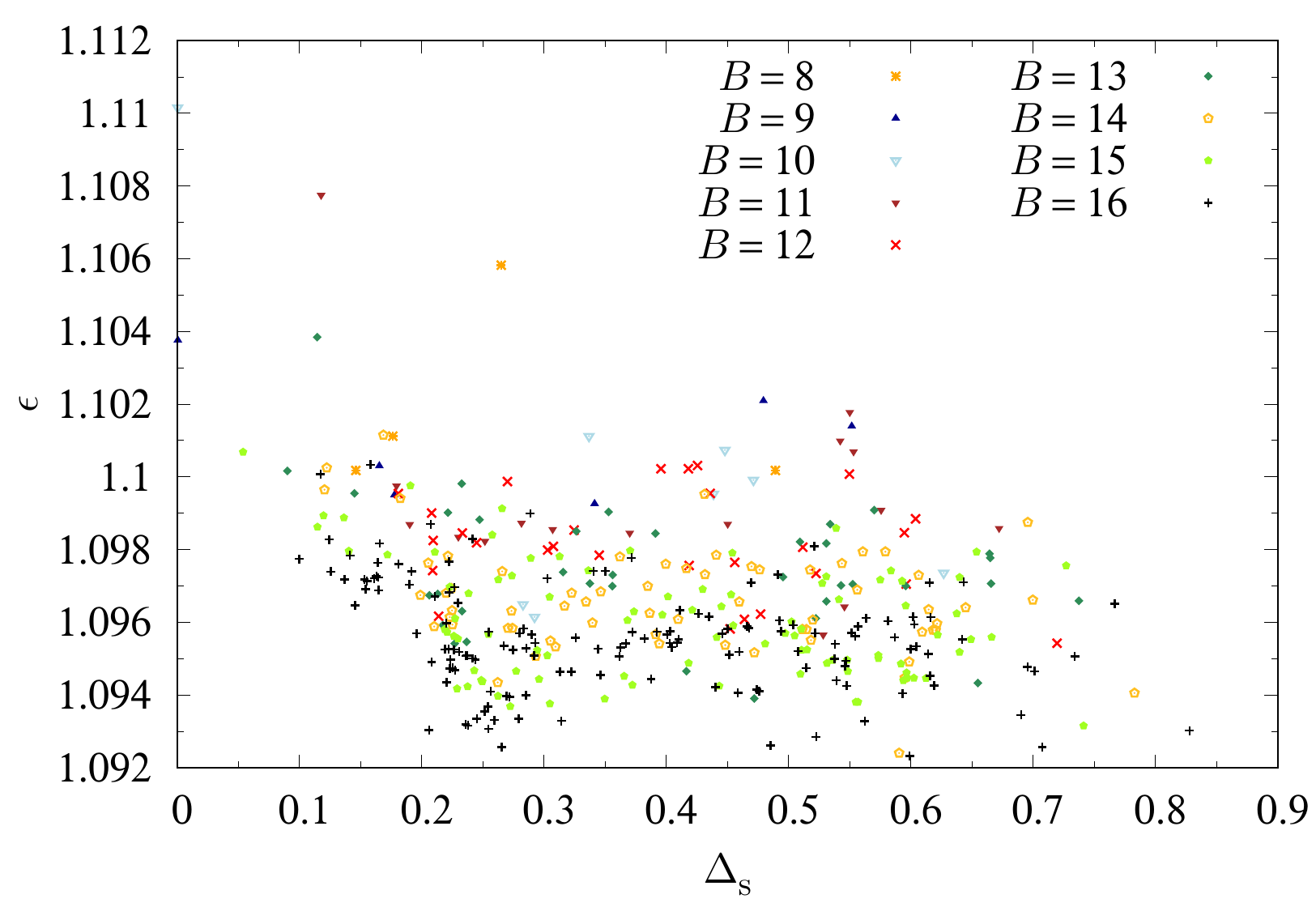}}}
\caption{ (a) Sphericity $\Delta_{\rm s}$ of the Skyrmions: the mean
and median as a box with the lines representing the spread. (b)
Sphericity $\Delta_{\rm s}$ of the $B=8$ through $16$ solutions versus
normalized energies $\epsilon$.}
\label{fig:stat_sphericity}
\end{center}
\end{figure}

\begin{figure}[!htp]
\begin{center}
\mbox{\subfloat[$B=4$ through $16$]{\includegraphics[width=0.33\linewidth]{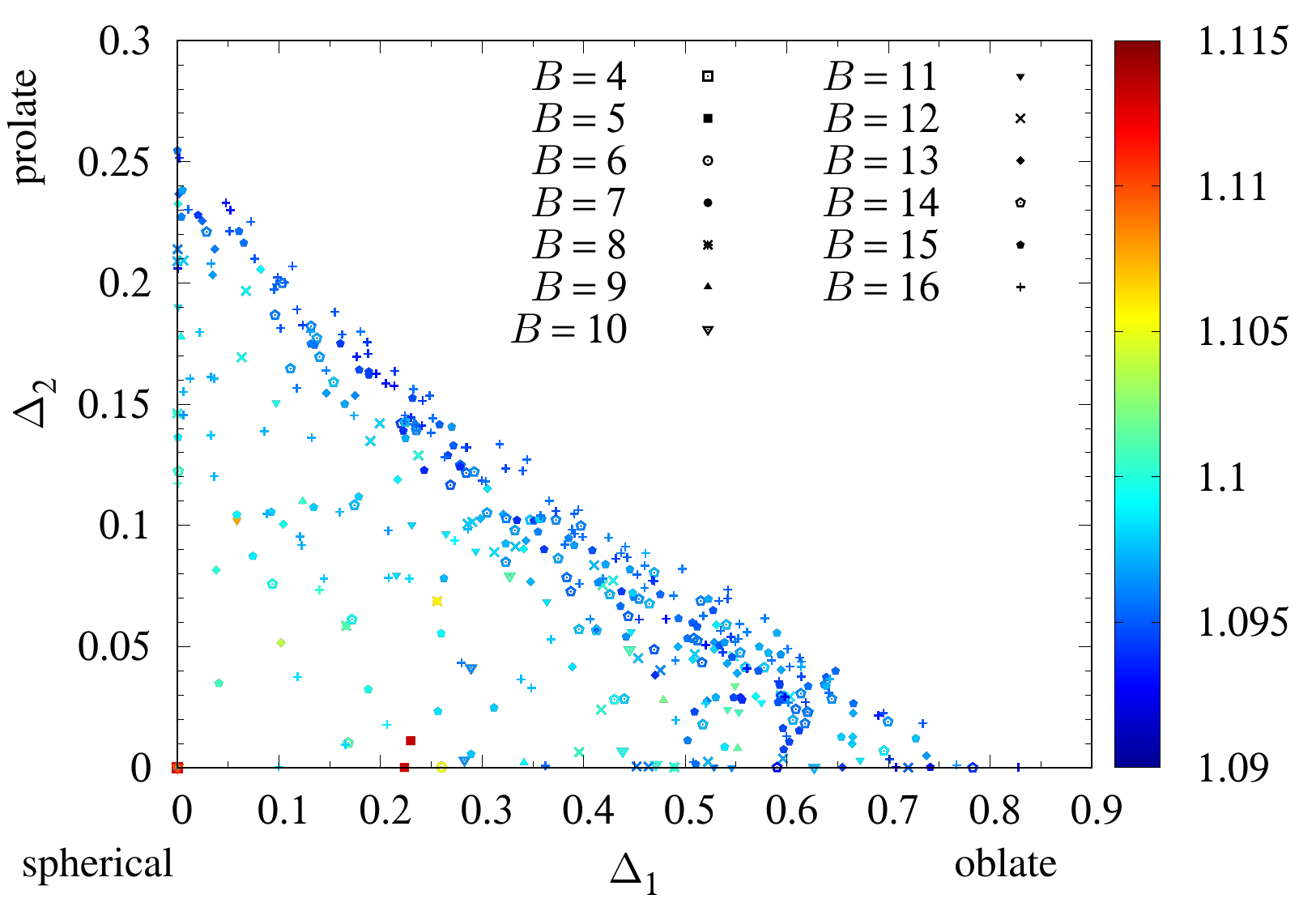}}
\subfloat[$B=8$]{\includegraphics[width=0.33\linewidth]{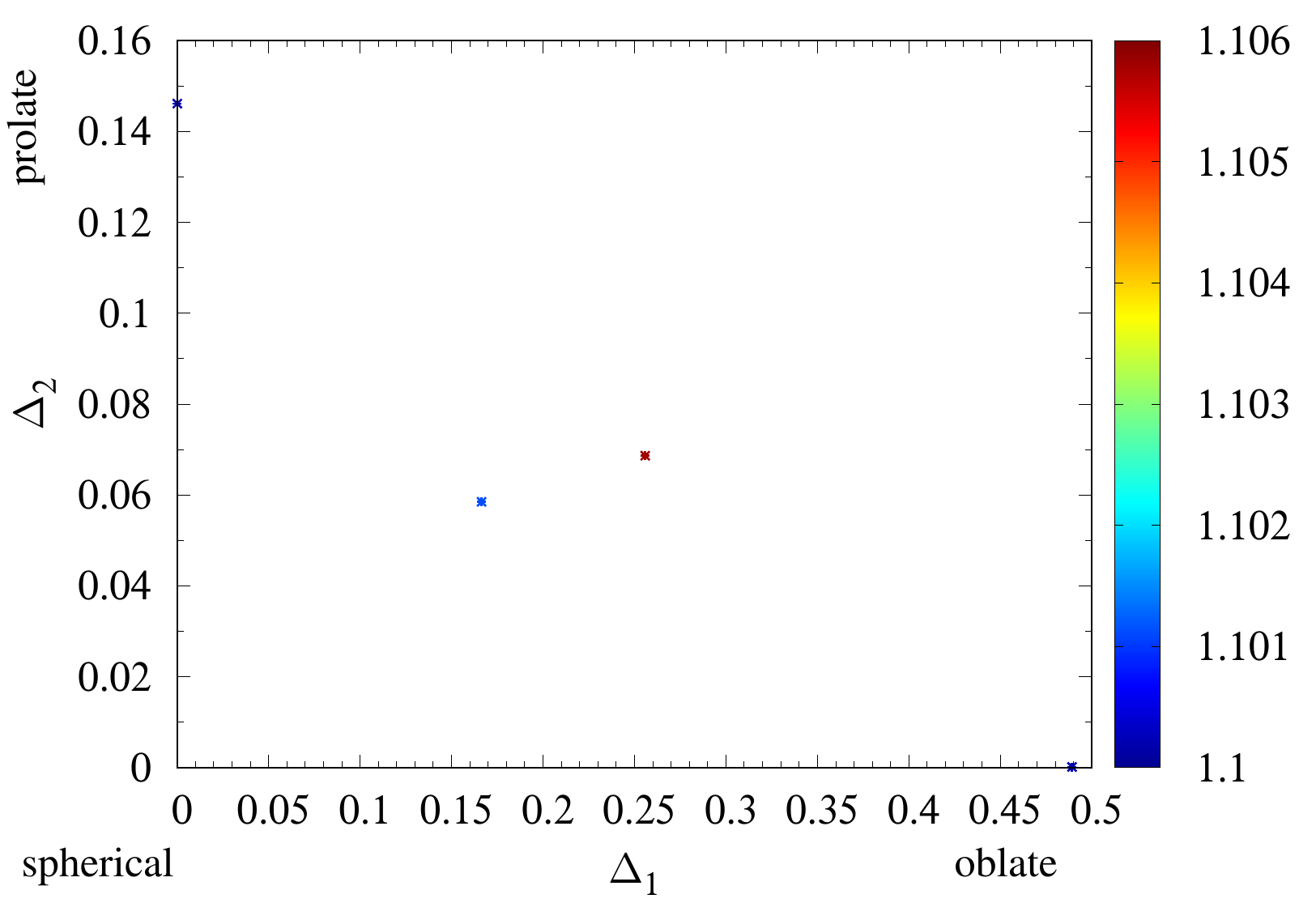}}
\subfloat[$B=9$]{\includegraphics[width=0.33\linewidth]{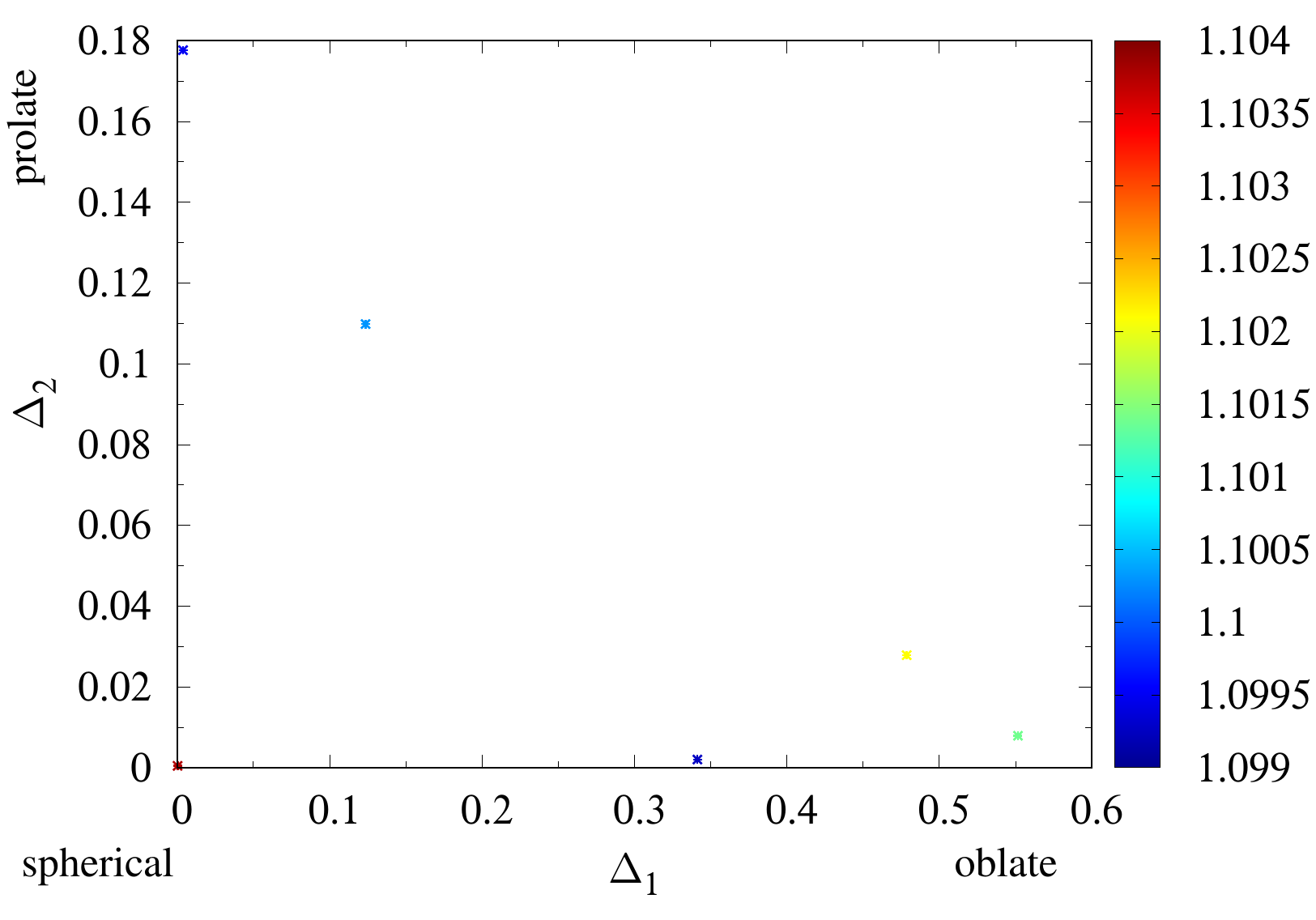}}}
\mbox{\subfloat[$B=10$]{\includegraphics[width=0.33\linewidth]{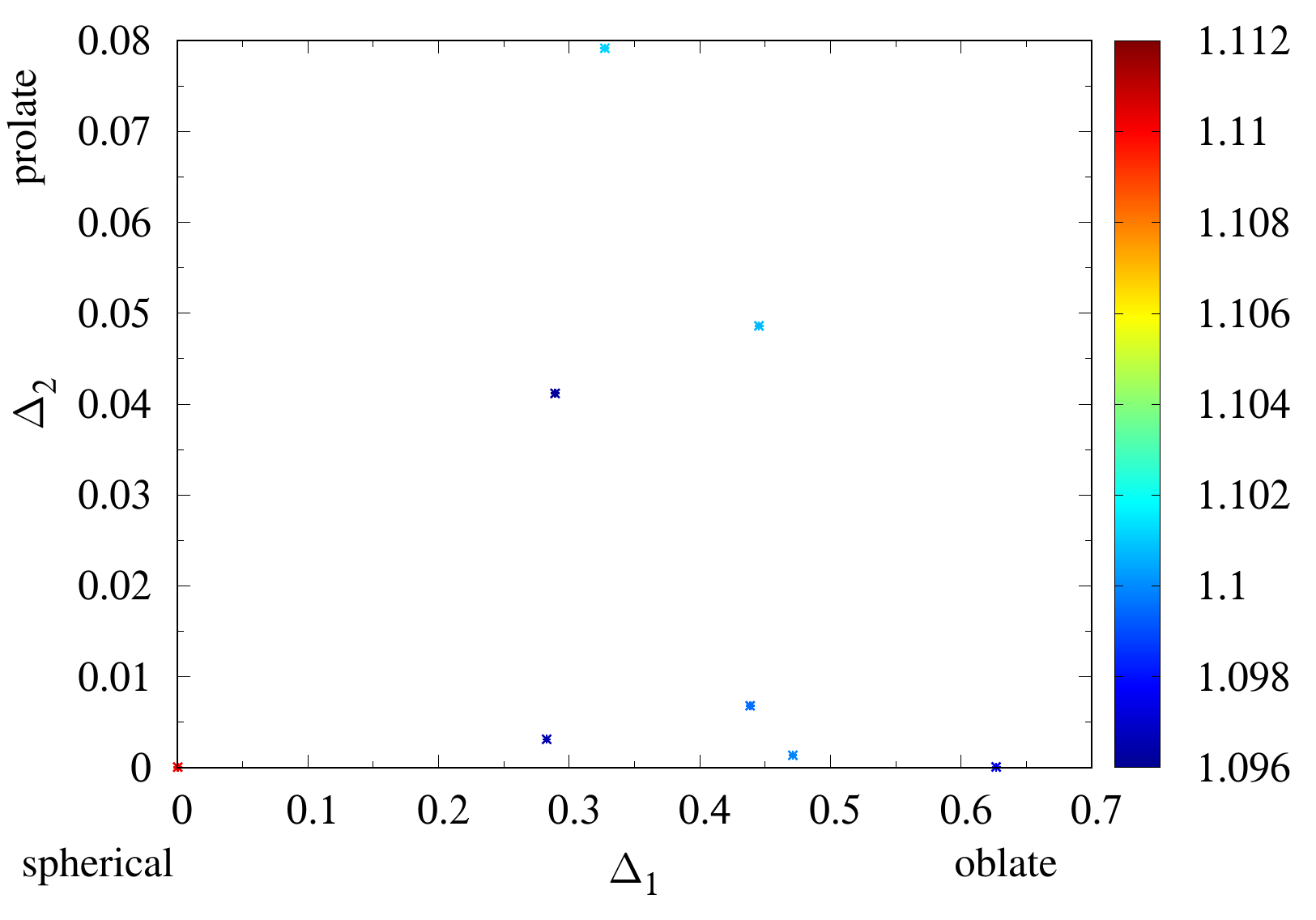}}
\subfloat[$B=11$]{\includegraphics[width=0.33\linewidth]{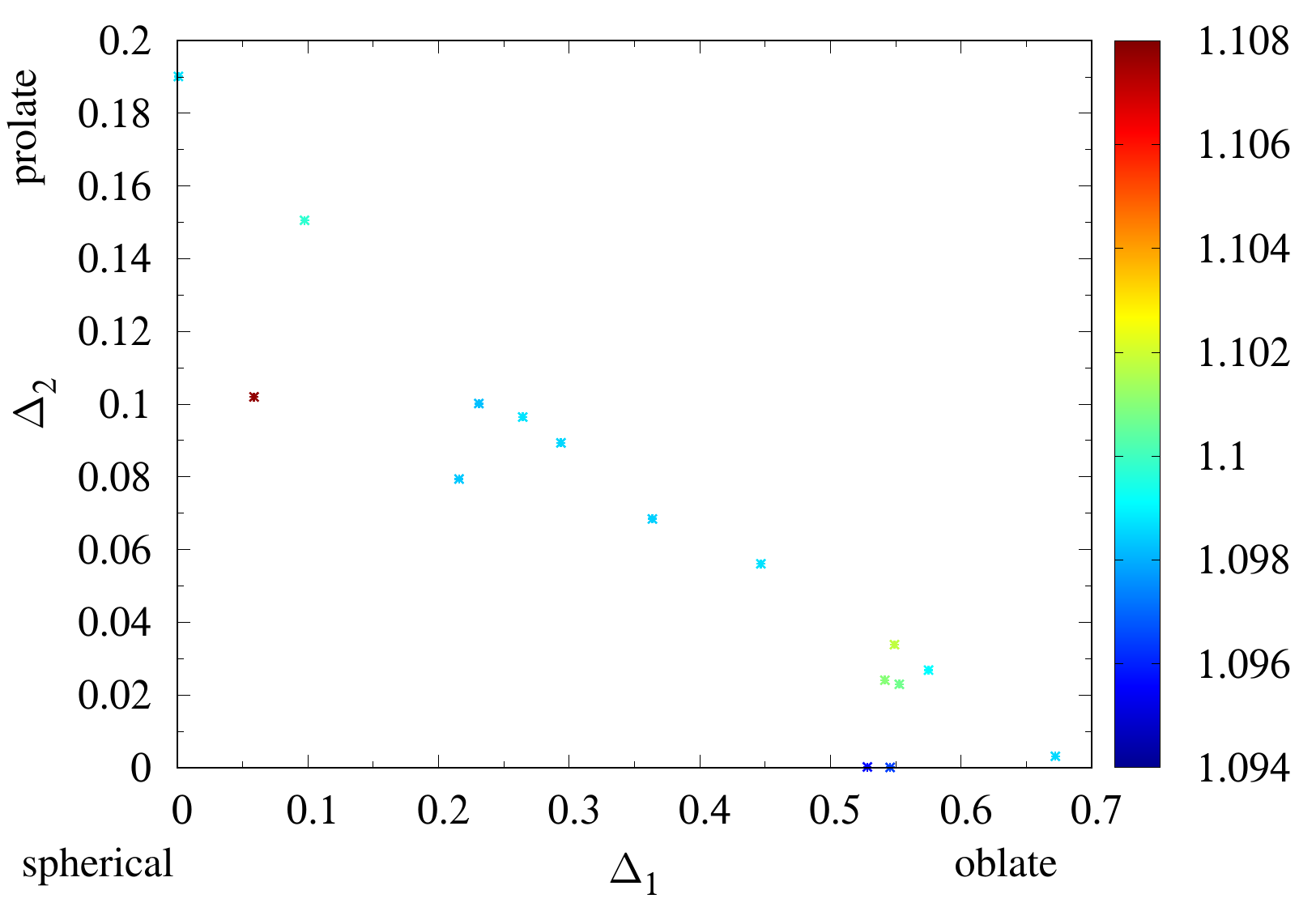}}
\subfloat[$B=12$]{\includegraphics[width=0.33\linewidth]{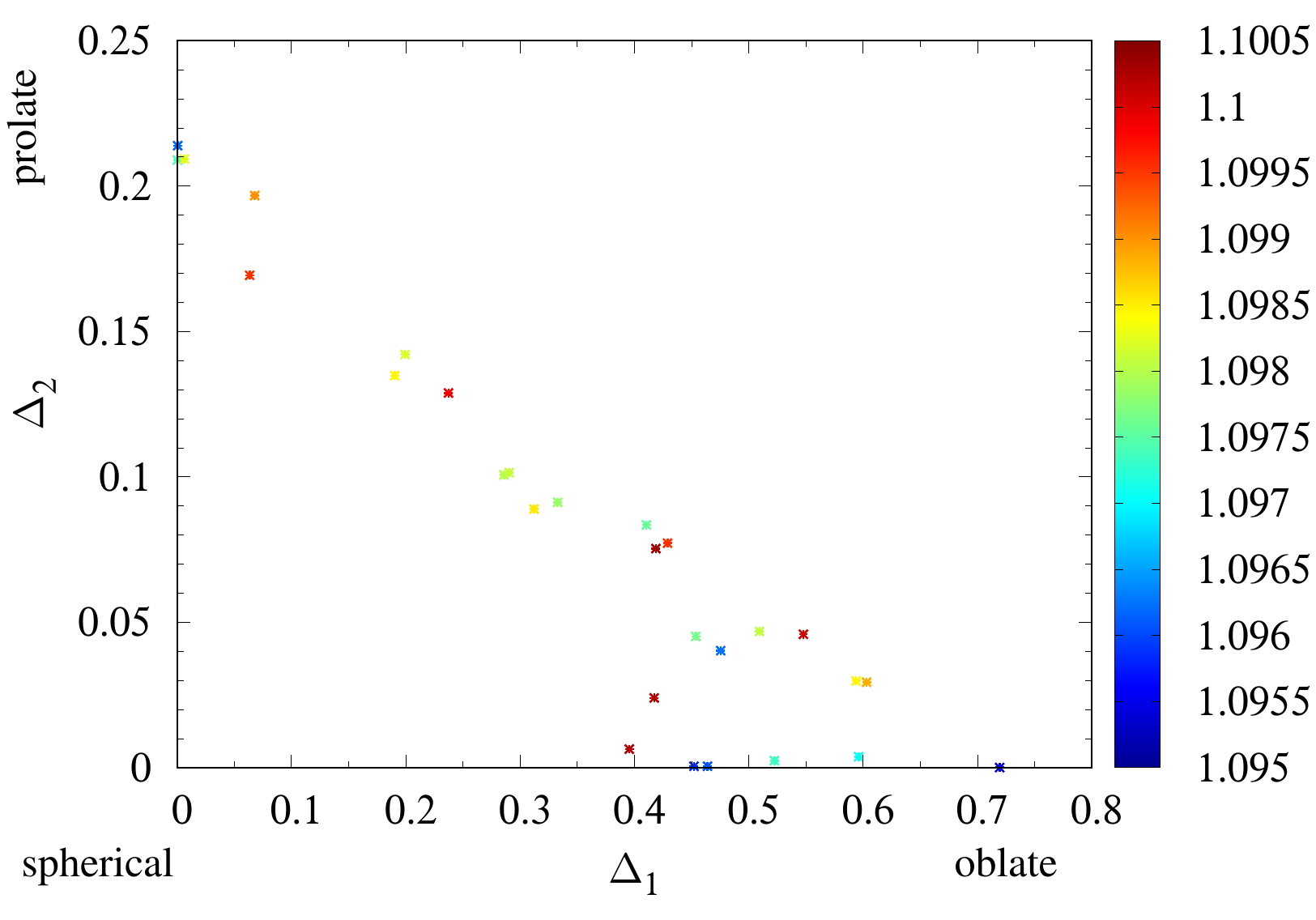}}}
\mbox{\subfloat[$B=13$]{\includegraphics[width=0.33\linewidth]{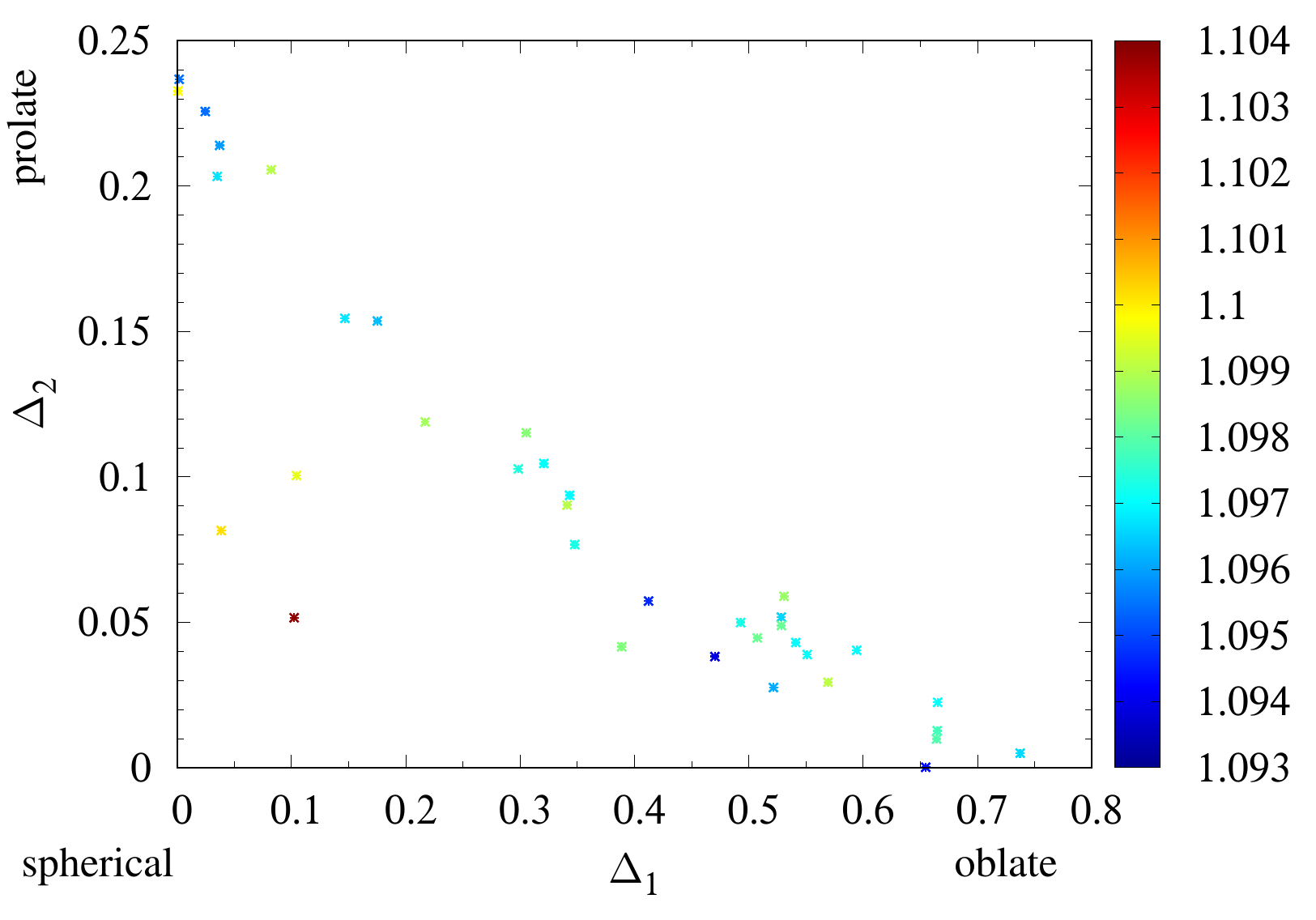}}
\subfloat[$B=14$]{\includegraphics[width=0.33\linewidth]{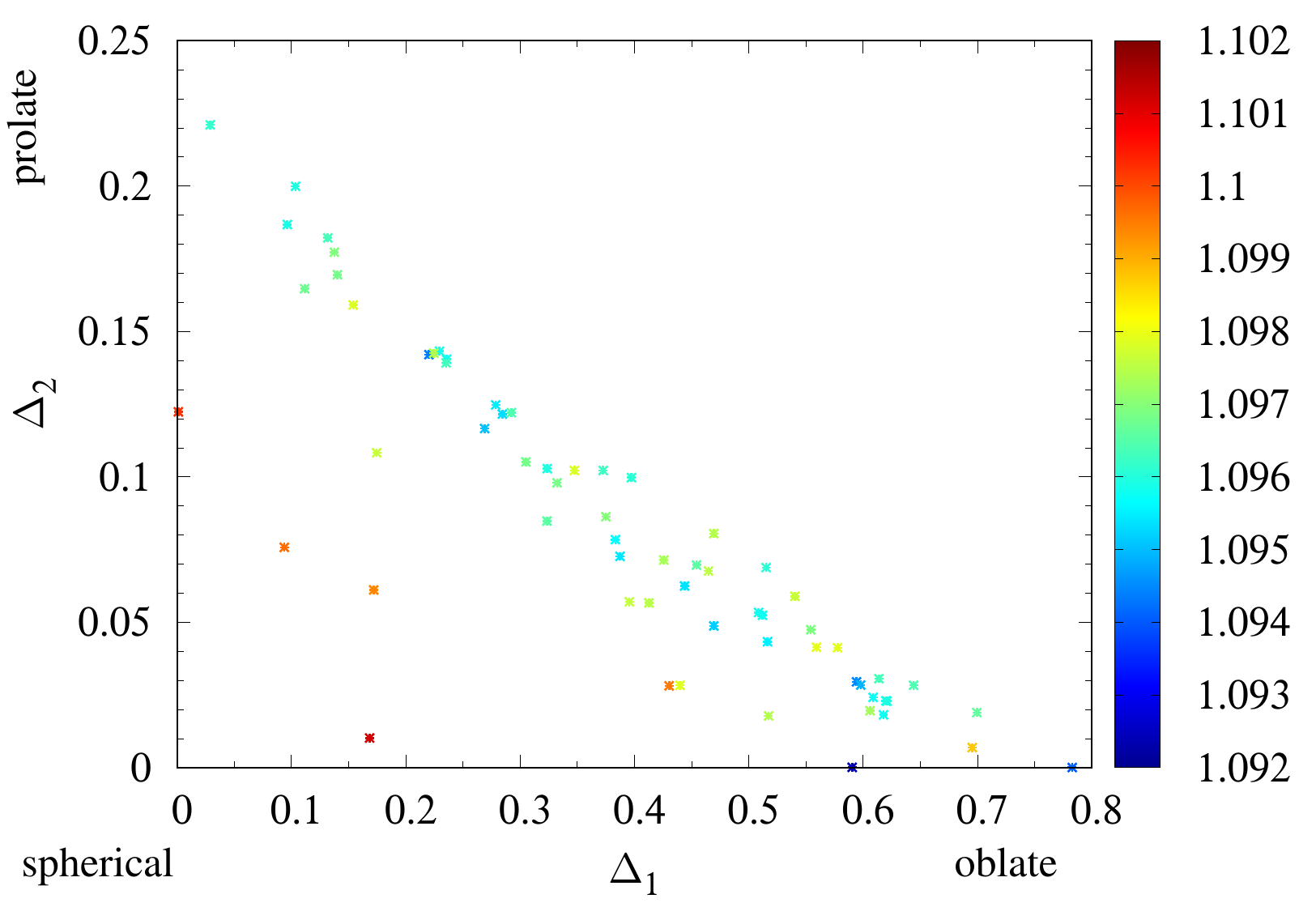}}
\subfloat[$B=15$]{\includegraphics[width=0.33\linewidth]{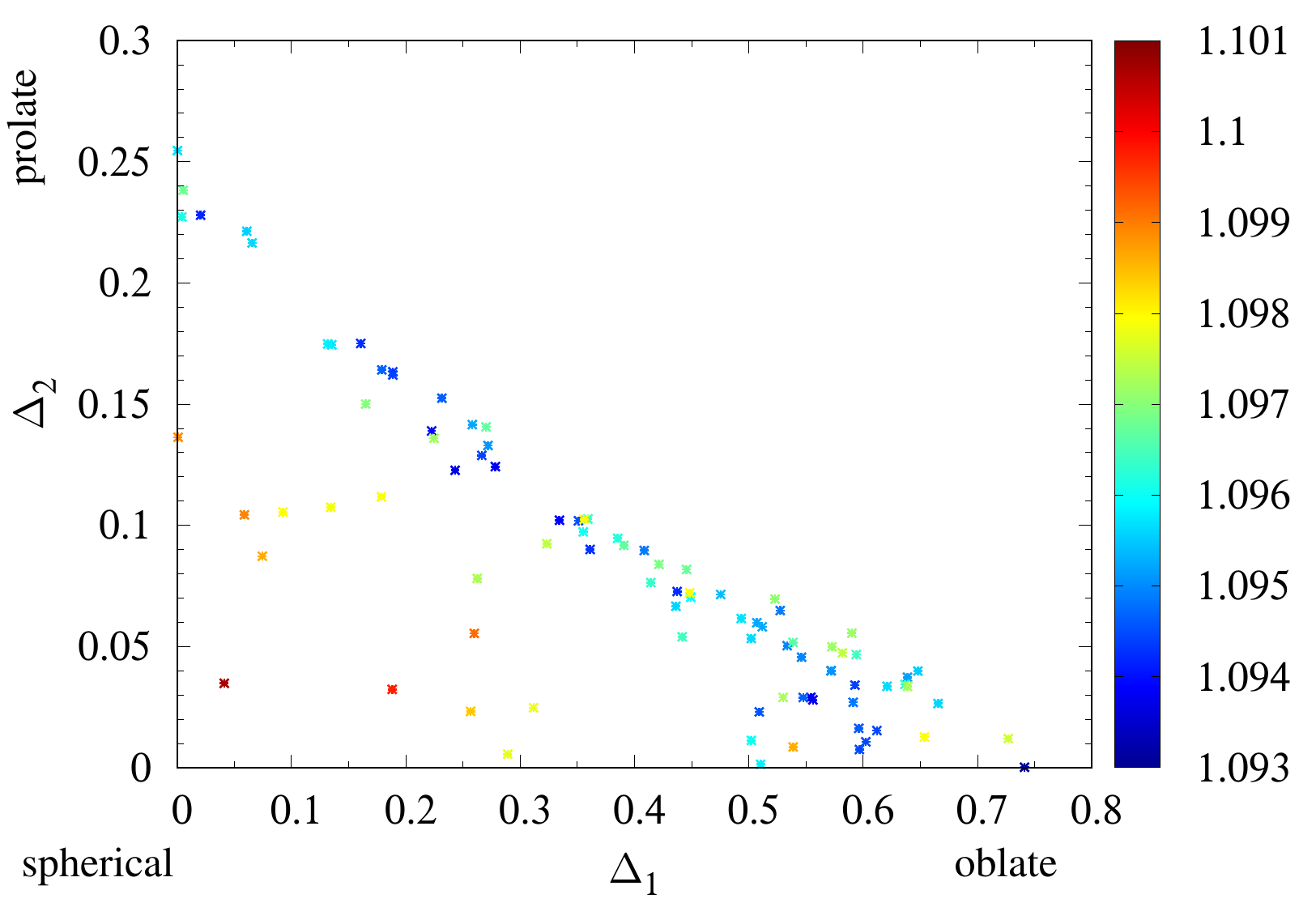}}}
\mbox{\subfloat[$B=16$]{\includegraphics[width=0.33\linewidth]{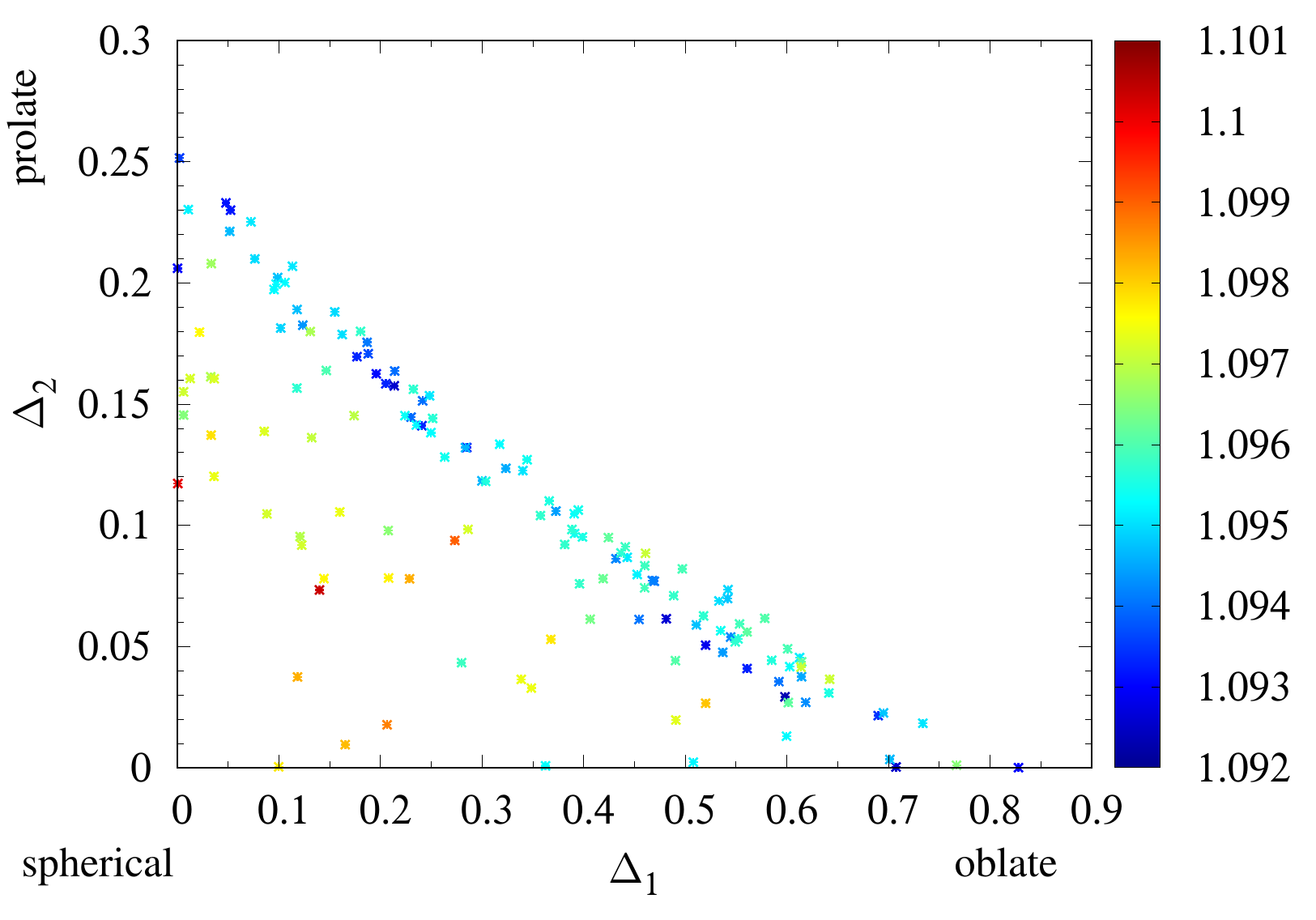}}}
\caption{Oblacity/Prolacity/Sphericity of solutions versus normalized energies
$\epsilon$. The color represents the energy: high and low energy
solutions are plotted in red and blue, respectively.} 
\label{fig:stat_obprolate}
\end{center}
\end{figure}

In fig.~\ref{fig:stat_sphericity}, we plot the sphericity $\Delta_{\rm s}$
versus the normalized energy $\epsilon$, defined as
\beq
\Delta_{\rm s} = \sqrt{\Delta_1^2 + \Delta_2^2},
\eeq
with
\beq
\Delta_1 = \frac{\lambda_2-\lambda_1}{\lambda_2+\lambda_1}, \qquad
\Delta_2 = \frac{\lambda_3-\lambda_2}{\lambda_3+\lambda_2}, 
\eeq
where $0<\lambda_1\leq\lambda_2\leq\lambda_3$ are the three
eigenvalues of the moment of inertia tensor $V$, see eq.~\eqref{eq:V},
with the eigenvalues $V_D$ found in eq.~\eqref{eq:VD} related as
$V_D=\diag(\lambda_1,\lambda_2,\lambda_3)$. 
Spherical symmetry implies $\Delta_1=\Delta_2=0$, an oblate
solution has $\Delta_2=0$ and $\Delta_1>0$, whereas a prolate solution
has $\Delta_1=0$ and $\Delta_2>0$. We see that the only low-energy
spherical Skyrmions are the small-$B$ fullerenes. There are
some high energy Skyrmions which are spherical, such as
the \skfigref{9}{skfig:B9c44}-
and \skfigref{10}{skfig:B10c11}-Skyrmions, but the small energy
solutions are not.

Finally, we show the relationship between sphericity, oblacity,
prolacity and normalized energy ($\epsilon$) in
fig.~\ref{fig:stat_obprolate}.
It is observed that the energy is generically minimized on a line
between oblate and prolate solutions and hence is generically not
spherically symmetric. However, there is no obvious energetic
advantage to being oblate over being prolate, or vice versa. These
results suggest that a spherical approximation, such as that made in
the RMA, is not a good starting point for finding
large $B$ Skyrmions with nonzero pion mass.

\section{Discussion and outlook}\label{sec:discussion}

In this paper, we have used large scale numerical computations on a
GPU cluster to find \numsols{} Skyrmion solutions, of which \newsols{}
are new. We find new families of Skyrmion such as different types of
graphene-like solutions, new chains and previously unstable
sub-clusters stabilized by adding some small Skyrmion.

We used a state-of-the-art arrested Newton flow algorithm implemented in
\texttt{CUDA C} and run on an \texttt{NVIDIA} GPU high-performance
computing cluster. 
The usage of GPU code, as opposed to traditional CPU code, was crucial
in order for us to be able to perform this large scale numerical
computation and finish within half a year.
For comparison, we tested a \texttt{C++}
implementation, i.e.~CPU code, of the same algorithm and estimate that the computation
with our current computing resources (a single \texttt{Intel Xeon} CPU
node with 24 cores) would have taken 45 years to
complete.\footnote{To be fair, we have used highly advanced memory
and optimization techniques to make the \texttt{CUDA C} code very
efficient, but we not made the same efforts for the \texttt{C++} code.}

We found that the new chains made from 2- and 3- tori sometimes had
lower energy than the usual cubic chains. Does this hold more
generally? The question can be answered for large $B$ by looking at
the infinite chains. Similarly
the \skfigref{10}{skfig:B10c1}-, \skfigref{13}{skfig:B13c26}-
and \skfigref{16}{skfig:B16c4}-Skyrmions are finite versions of an
infinite slab and could be studied in this limit. Other graphene
solutions are also cut from the infinite two-layer hexagonal
lattice. Hence their properties may be deduced from a careful study of
the Skyrmion lattice. There may even be a way to generate certain
shapes from the lattice in a (semi-)analytic way.

For rigid body quantization to be a good approximation, the quantized
Skyrmion should be isolated in 
configuration space. The wavefunction is concentrated around a 
single Skyrmion, not overlapping with any other solutions.
This should work well for $B<8$, where there are few
solutions. Indeed, with an appropriate 
calibration of the Skyrme model, rigid body
quantization does give reasonably accurate nuclear spectra, albeit
only for even baryon numbers \cite{Manko:2007pr}.
For odd $B$, additional ingredients in the quantization become
necessary in order to produce realistic spectra. 
There are extra modes that must be quantized, in the
spirit of the semi-classical expansion as a low-energy effective
field theory, as light modes -- vibrational
modes \cite{Gudnason:2018bju} -- and turn out to be crucial for the
nuclear spectrum of e.g.~Lithium-7 \cite{Halcrow:2015rvz}. Again, relying on
these light modes only works on the assumption that the Skyrmion is
isolated in configuration space. An obvious laborious task for future
work would be to calculate the vibrational spectra of all the newly
found Skyrmion solutions and further more use this information to
construct nuclear spectra.

The sheer number of classical solutions show that the Skyrmion landscape is
incredibly complicated -- and the minima are only the starting
point. Morse theory guarantees that any two minima have a saddle point
between them. So the landscape is likely rather flat, with many low
energy paths between the many low energy solutions. This picture is in
stark contrast to the assumptions of rigid body quantization.
Our results suggest that a new paradigm for Skyrmion quantization is
required: one where we take 
the enormous family of low energy configurations seriously. How this
is done in a computationally efficient way is unknown. Perhaps our
results can be used as direct input in some new method. Another option
is to use ADHM data to produce large families of Skyrme
configurations \cite{Cork:2021uov}.

Further future avenues of research include
altering the Skyrme model to yield smaller classical binding energies,
see
e.g.~\cite{Sutcliffe:2010et,Sutcliffe:2011ig,Adam:2010fg,Adam:2010ds,Gillard:2015eia,Gudnason:2016mms,Gudnason:2018jia,Naya:2018kyi,Gudnason:2020arj},
and recalculating all the classical solutions in such cases.
We expect that there will be many more solutions in a weakly bound
Skyrme model with realistic binding energies \cite{Gillard:2015eia},
which would further limit the practical usefulness of the
semi-classical quantization.

Our method to generate configurations is simple: we generate $B$
random 1-Skyrmions and place them near one another. This naive method
has produced the largest number of Skyrmions ever found. Perhaps the
method could be implemented in other soliton systems to find new
solutions. 

We hope that our work will serve as motivation for the invention of
new methods to analytically describe classical Skyrmions and provoke a
complete rethink of Skyrmion quantization.

\subsection*{Acknowledgments}

We thank Stefano Bolognesi for correspondence.
We also thank the academic
atmosphere at the solitons@work network 
(\href{http://solitonsatwork.net}{solitonsatwork.net}), which
motivated us to carry out this research. 
S.~B.~G.~thanks the Outstanding Talent Program of Henan University for
partial support.
The work of S.~B.~G.~is supported by the National Natural Science
Foundation of China (Grants No.~11675223 and No.~12071111).
C.~H.~is supported by the University of Leeds as an
Academic Development Fellow.

\phantomsection
\addcontentsline{toc}{section}{References}
\bibliographystyle{utphys}
\bibliography{refs}

\newpage
\pagestyle{fancy}
\fancyhead{}
\fancyhead[RO]{\ref{B:1}, $\ldots$ \ref{B:5}, $\ldots$ \ref{B:8}, $\ldots$ \ref{B:10}, \ref{B:11}, \ref{B:12}, \ref{B:13}, \ref{B:14}, \ref{B:15}, and \ref{B:16}}
\fancyhead[LO]{\it A Sm\"org\aa sbord of Skyrmions:}
\fancyfoot{}
\fancyfoot[CO]{-- \thepage\ --}
\appendix
\section{Numerical solutions}\label{app:sols}

\def\figheight{85pt}
\refstepcounter{Bnum}\label{B:1}
\begin{longtable}{|lc|r|l|}\hline
\multicolumn{4}{|c|}{$B=1$\ (1 solution)}\\*\hline\hline

1$_{{\rm \skfig\label{skfig:B1c1}}}$ &
\multirow{6}{*}{\includegraphics[height=\figheight]{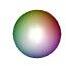}} & 
$\epsilon$ & 1.2171 $\pm$ 0.0002 \\*
&& $b$ & 0.9998 \\*
&& $\mathfrak{r}$ & 0.7761 \\*
&& $V_D$ & (47.4,47.4,47.4) \\*
&& $U_D$ & (47.5,47.5,47.5) \\*
100\% && sym & $O(3)$ \cite{Skyrme:1961vq,Skyrme:1962vh,Battye:2000se} \\*
\cline{1-4}\multicolumn{4}{|p{0.97\linewidth}|}{The spherically symmetric Skyrmion.} \\
\hline
\end{longtable}

\refstepcounter{Bnum}\label{B:2}
\begin{longtable}{|lc|r|l|}\hline
\multicolumn{4}{|c|}{$B=2$\ (1 solution)}\\*\hline\hline

2$_{{\rm \skfig\label{skfig:B2c1}}}$ &
\multirow{6}{*}{\includegraphics[height=\figheight]{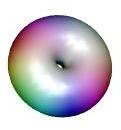}} & 
$\epsilon$ & 1.1690 $\pm$ 0.0001 \\*
&& $b$ & 0.9999 \\*
&& $\mathfrak{r}$ & 1.0550 \\*
&& $V_D$ & (153.7,154.0,275.3) \\*
&& $U_D$ & (68.9,97.1,97.1) \\*
100\% && sym & $D_{{\infty}h}$ \cite{Verbaarschot:1987au,Battye:2000se} \\*
\cline{1-4}\multicolumn{4}{|p{0.97\linewidth}|}{The only Skyrmion with toroidal symmetry. We will later refer to this solution as ``the 2-torus".} \\
\hline
\end{longtable}

\refstepcounter{Bnum}\label{B:3}
\begin{longtable}{|lc|r|l|}\hline
\multicolumn{4}{|c|}{$B=3$\ (1 solution)}\\*\hline\hline

3$_{{\rm \skfig\label{skfig:B3c1}}}$ &
\multirow{6}{*}{\includegraphics[height=\figheight]{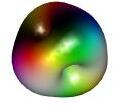}} & 
$\epsilon$ & 1.1378 $\pm$ 0.0001 \\*
&& $b$ & 0.9999 \\*
&& $\mathfrak{r}$ & 1.2244 \\*
&& $V_D$ & (402.2,402.5,402.6) \\*
&& $U_D$ & (124.0,124.0,124.0) \\*
100\% && sym & $T_d$ \cite{Braaten:1989rg} \\*
\cline{1-4}\multicolumn{4}{|p{0.97\linewidth}|}{The energy density is focused on the edges of a tetrahedron. We will later refer to this solution as ``the tetrahedron".} \\
\hline
\end{longtable}

\refstepcounter{Bnum}\label{B:4}
\begin{longtable}{|lc|r|l|}\hline
\multicolumn{4}{|c|}{$B=4$\ (1 solution)}\\*\hline\hline

4$_{{\rm \skfig\label{skfig:B4c2}}}$ &
\multirow{6}{*}{\includegraphics[height=\figheight]{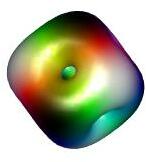}} & 
$\epsilon$ & 1.1129 $\pm$ 0.0001 \\*
&& $b$ & 0.9999 \\*
&& $\mathfrak{r}$ & 1.3588 \\*
&& $V_D$ & (666.1,666.3,666.4) \\*
&& $U_D$ & (177.2,148.0,148.0) \\*
100\% && sym & $O_h$ \cite{Braaten:1989rg} \\*
\cline{1-4}\multicolumn{4}{|p{0.97\linewidth}|}{The energy density is focused on the edges of a cube. We will later refer to this solution as ``the cube".} \\
\hline
\end{longtable}

\newpage
\refstepcounter{Bnum}\label{B:5}
\begin{longtable}{|lc|r|l|}\hline
\multicolumn{4}{|c|}{$B=5$\ (2 solutions)}\\*\hline\hline

5$_{{\rm \skfig\label{skfig:B5c1}}}$ &
\multirow{6}{*}{\includegraphics[height=\figheight]{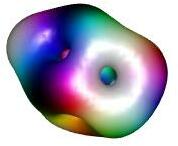}} & 
$\epsilon$ & 1.1131 $\pm$ 0.0001 \\*
&& $b$ & 0.9999 \\*
&& $\mathfrak{r}$ & 1.5162 \\*
&& $V_D$ & (775.9,1222.7,1223.0) \\*
&& $U_D$ & (180.9,201.3,201.3) \\*
98.4\% && sym & $D_{2d}$ \cite{Braaten:1989rg} \\*
\cline{1-4}\multicolumn{4}{|p{0.97\linewidth}|}{Dihedrally symmetric; can be interpreted as the cube with a single Skyrmion attached at an edge.} \\
\hline

5$_{{\rm \skfig\label{skfig:B5c5}}}$ &
\multirow{6}{*}{\includegraphics[height=\figheight]{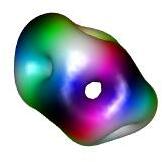}} & 
$\epsilon$ & 1.1132 $\pm$ 0.0001 \\*
&& $b$ & 0.9999 \\*
&& $\mathfrak{r}$ & 1.5229 \\*
&& $V_D$ & (769.6,1228.9,1256.7) \\*
&& $U_D$ & (181.6,205.2,198.6) \\*
1.63\% && sym & $C_{2v}$ \\*
\cline{1-4}\multicolumn{4}{|p{0.97\linewidth}|}{Fours holes on top and bottom, out-of-phase. The solution is a deformation of the $D_{4d}$ symmetric solution considered in ref.~\cite{Manko:2007pr}.} \\
\hline
\end{longtable}

\refstepcounter{Bnum}\label{B:6}
\begin{longtable}{|lc|r|l|}\hline
\multicolumn{4}{|c|}{$B=6$\ (1 solution)}\\*\hline\hline

6$_{{\rm \skfig\label{skfig:B6c1}}}$ &
\multirow{6}{*}{\includegraphics[height=\figheight]{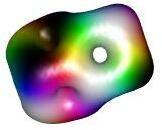}} & 
$\epsilon$ & 1.1051 $\pm$ 0.0001 \\*
&& $b$ & 0.9999 \\*
&& $\mathfrak{r}$ & 1.6518 \\*
&& $V_D$ & (1054.6,1796.4,1797.3) \\*
&& $U_D$ & (225.0,231.9,232.1) \\*
100\% && sym & $D_{4d}$ \cite{Battye:1997qq} \\*
\cline{1-4}\multicolumn{4}{|p{0.97\linewidth}|}{Dihedrally symmetric. Can be thought of as three 2-tori stacked, or the cube with a 2-torus attached to a face.} \\
\hline
\end{longtable}

\refstepcounter{Bnum}\label{B:7}
\begin{longtable}{|lc|r|l|}\hline
\multicolumn{4}{|c|}{$B=7$\ (1 solution)}\\*\hline\hline

7$_{{\rm \skfig\label{skfig:B7c1}}}$ &
\multirow{6}{*}{\includegraphics[height=\figheight]{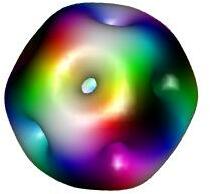}} & 
$\epsilon$ & 1.0958 $\pm$ 0.0001 \\*
&& $b$ & 0.9999 \\*
&& $\mathfrak{r}$ & 1.6952 \\*
&& $V_D$ & (1887.5,1887.7,1888.1) \\*
&& $U_D$ & (257.4,257.4,257.4) \\*
100\% && sym & $Y_h$ \cite{Battye:1997qq} \\*
\cline{1-4}\multicolumn{4}{|p{0.97\linewidth}|}{The icosahedrally symmetric Skyrmion.} \\
\hline
\end{longtable}

\newpage
\refstepcounter{Bnum}\label{B:8}
\begin{longtable}{|lc|r|l|}\hline
\multicolumn{4}{|c|}{$B=8$\ (4 solutions)}\\*\hline\hline

8$_{{\rm \skfig\label{skfig:B8c1}}}$ &
\multirow{6}{*}{\includegraphics[height=\figheight]{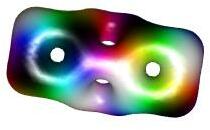}} & 
$\epsilon$ & 1.1002 $\pm$ 0.0001 \\*
&& $b$ & 0.9999 \\*
&& $\mathfrak{r}$ & 2.0159 \\*
&& $V_D$ & (1387.6,4045.7,4047.0) \\*
&& $U_D$ & (325.9,298.6,291.1) \\*
32.4\% && sym & $D_{4h}$ \cite{Battye:2006na} \\*
\cline{1-4}\multicolumn{4}{|p{0.97\linewidth}|}{Two cubes, one rotated by 90 degrees around the axis joining them.} \\
\hline
\end{longtable}

\refstepcounter{Bnum}\label{B:9}
\begin{longtable}{|lc|r|l|}\hline
\multicolumn{4}{|c|}{$B=9$\ (6 solutions)}\\*\hline\hline

9$_{{\rm \skfig\label{skfig:B9c28}}}$ &
\multirow{6}{*}{\includegraphics[height=\figheight]{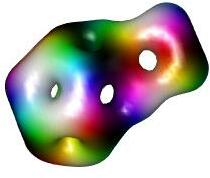}} & 
$\epsilon$ & 1.0993 $\pm$ 0.0001 \\*
&& $b$ & 0.9999 \\*
&& $\mathfrak{r}$ & 2.0123 \\*
&& $V_D$ & (2091.6,4258.9,4276.3) \\*
&& $U_D$ & (348.9,326.6,329.8) \\*
55.0\% && sym & $C_{1h}$ \\*
\cline{1-4}\multicolumn{4}{|p{0.97\linewidth}|}{The \skfigref{7}{skfig:B7c1}-Skyrmion with a 2-torus attached to a face.} \\
\hline
\end{longtable}

\newpage
\refstepcounter{Bnum}\label{B:10}
\begin{longtable}{|lc|r|l|}\hline
\multicolumn{4}{|c|}{$B=10$\ (8 solutions)}\\*\hline\hline

10$_{{\rm \skfig\label{skfig:B10c1}}}$ &
\multirow{6}{*}{\includegraphics[height=\figheight]{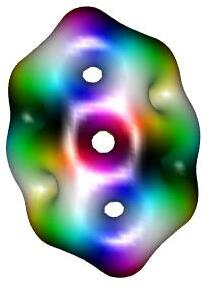}} & 
$\epsilon$ & 1.0961 $\pm$ 0.0001 \\*
&& $b$ & 0.9999 \\*
&& $\mathfrak{r}$ & 2.0606 \\*
&& $V_D$ & (2579.7,4679.9,5081.8) \\*
&& $U_D$ & (356.0,371.9,369.9) \\*
83.6\% && sym & $D_{2h}$ \cite{Battye:2006tb} \\*
\cline{1-4}\multicolumn{4}{|p{0.97\linewidth}|}{Can be interpreted as symmetric graphene or a collapsed fullerene.} \\
\hline
\end{longtable}

\refstepcounter{Bnum}\label{B:11}
\begin{longtable}{|lc|r|l|}\hline
\multicolumn{4}{|c|}{$B=11$\ (16 solutions)}\\*\hline\hline

11$_{{\rm \skfig\label{skfig:B11c6}}}$ &
\multirow{6}{*}{\includegraphics[height=\figheight]{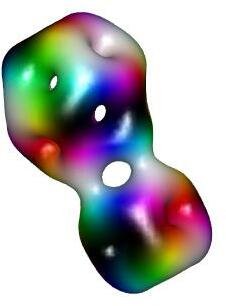}} & 
$\epsilon$ & 1.0957 $\pm$ 0.0001 \\*
&& $b$ & 0.9999 \\*
&& $\mathfrak{r}$ & 2.3915 \\*
&& $V_D$ & (2475.2,8018.3,8021.6) \\*
&& $U_D$ & (421.5,402.7,405.6) \\*
33.4\% && sym & $C_{1h}$ \\*
\cline{1-4}\multicolumn{4}{|p{0.97\linewidth}|}{The \skfigref{7}{skfig:B7c1}-Skyrmion with the cube attached to a face. The reflection plane of the cube and icosahedron match.} \\
\hline
\end{longtable}

\newpage
\refstepcounter{Bnum}\label{B:12}
\begin{longtable}{|lc|r|l|}\hline
\multicolumn{4}{|c|}{$B=12$\ (28 solutions)}\\*\hline\hline

12$_{{\rm \skfig\label{skfig:B12c51}}}$ &
\multirow{6}{*}{\includegraphics[height=\figheight]{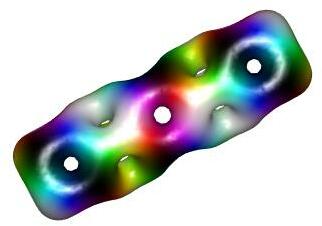}} & 
$\epsilon$ & 1.0954 $\pm$ 0.0001 \\*
&& $b$ & 0.9999 \\*
&& $\mathfrak{r}$ & 2.7878 \\*
&& $V_D$ & (2092.9,12829.9,12831.4) \\*
&& $U_D$ & (473.5,438.6,448.4) \\*
0.80\% && sym & $D_{4h}$ \cite{Battye:2006na} \\*
\cline{1-4}\multicolumn{4}{|p{0.97\linewidth}|}{A slice of the infinite twisted chain. Three cubes with the middle cube twisted by 90 degrees around the axis that joins the cubes.} \\
\hline
\end{longtable}

\refstepcounter{Bnum}\label{B:13}
\begin{longtable}{|lc|r|l|}\hline
\multicolumn{4}{|c|}{$B=13$\ (36 solutions)}\\*\hline\hline

13$_{{\rm \skfig\label{skfig:B13c26}}}$ &
\multirow{6}{*}{\includegraphics[height=\figheight]{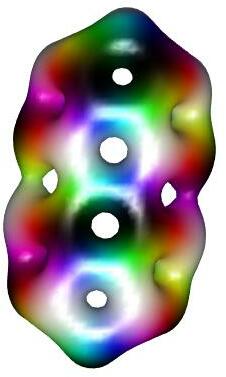}} & 
$\epsilon$ & 1.0939 $\pm$ 0.0001 \\*
&& $b$ & 0.9999 \\*
&& $\mathfrak{r}$ & 2.4740 \\*
&& $V_D$ & (3455.2,9592.6,10354.2) \\*
&& $U_D$ & (467.5,486.7,480.4) \\*
17.6\% && sym & $D_{2h}$ \\*
\cline{1-4}\multicolumn{4}{|p{0.97\linewidth}|}{Symmetric graphene.} \\
\hline
\end{longtable}

\refstepcounter{Bnum}\label{B:14}
\begin{longtable}{|lc|r|l|}\hline
\multicolumn{4}{|c|}{$B=14$\ (64 solutions)}\\*\hline\hline

14$_{{\rm \skfig\label{skfig:B14c46}}}$ &
\multirow{6}{*}{\includegraphics[height=\figheight]{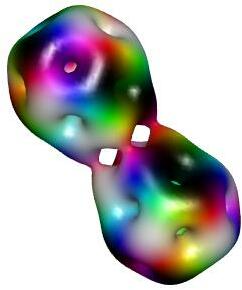}} & 
$\epsilon$ & 1.0924 $\pm$ 0.0001 \\*
&& $b$ & 0.9999 \\*
&& $\mathfrak{r}$ & 2.7802 \\*
&& $V_D$ & (3642.5,14143.3,14146.6) \\*
&& $U_D$ & (517.2,510.7,509.9) \\*
5.30\% && sym & $D_{5d}$ \\*
\cline{1-4}\multicolumn{4}{|p{0.97\linewidth}|}{Two \skfigref{7}{skfig:B7c1}-Skyrmions attached to one another symmetrically.} \\
\hline
\end{longtable}

\refstepcounter{Bnum}\label{B:15}
\begin{longtable}{|lc|r|l|}\hline
\multicolumn{4}{|c|}{$B=15$\ (92 solutions)}\\*\hline\hline

15$_{{\rm \skfig\label{skfig:B15c6}}}$ &
\multirow{6}{*}{\includegraphics[height=\figheight]{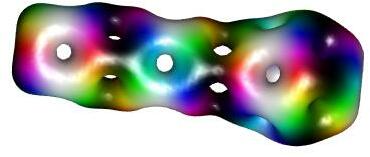}} & 
$\epsilon$ & 1.0932 $\pm$ 0.0001 \\*
&& $b$ & 0.9999 \\*
&& $\mathfrak{r}$ & 3.2074 \\*
&& $V_D$ & (3180.0,21397.4,21404.2) \\*
&& $U_D$ & (569.3,552.0,552.4) \\*
0.52\% && sym & $C_{1h}$ \\*
\cline{1-4}\multicolumn{4}{|p{0.97\linewidth}|}{A \skfigref{7}{skfig:B7c1}-Skyrmion attached to a \skfigref{8}{skfig:B8c1}-Skyrmion.} \\
\hline
\end{longtable}

\refstepcounter{Bnum}\label{B:16}
\begin{longtable}{|lc|r|l|}\hline
\multicolumn{4}{|c|}{$B=16$\ (147 solutions)}\\*\hline\hline

16$_{{\rm \skfig\label{skfig:B16c4}}}$ &
\multirow{6}{*}{\includegraphics[height=\figheight]{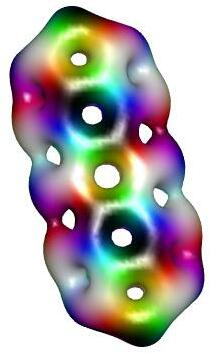}} & 
$\epsilon$ & 1.0923 $\pm$ 0.0001 \\*
&& $b$ & 0.9999 \\*
&& $\mathfrak{r}$ & 2.9037 \\*
&& $V_D$ & (4345.0,17280.1,18320.6) \\*
&& $U_D$ & (576.5,584.6,600.0) \\*
3.82\% && sym & $D_{2h}$ \\*
\cline{1-4}\multicolumn{4}{|p{0.97\linewidth}|}{Symmetric graphene.} \\
\hline
\end{longtable}

\end{document}